\documentclass[journal]{IEEEtran}
\usepackage{graphicx}
\usepackage{cite}
\usepackage{picinpar}
\usepackage{amsmath,amsfonts,amssymb,amsthm}
\usepackage{url}
\usepackage{flushend}
\usepackage{soul}
\usepackage{multirow}
\usepackage{pifont}
\usepackage{alltt}
\usepackage[hidelinks]{hyperref}
\usepackage{enumerate}
\usepackage{siunitx}
\usepackage{epstopdf}
\usepackage{pbox}
\usepackage{booktabs}
\usepackage{xcolor}

\DeclareMathOperator{\argmin}{arg\,min\,}
\DeclareMathOperator{\tr}{\mathrm{tr}}
\DeclareMathOperator{\mean}{\mathbb{E}}

\begin{document}
\title{Sigma-point Kalman Filter with Nonlinear Unknown Input Estimation via Optimization and Data-driven Approach for Dynamic Systems}

\author{Junn Yong Loo$^{1,2}$, Ze Yang Ding$^{1}$, Vishnu Monn Baskaran$^{2}$, Surya Girinatha Nurzaman$^{1}$, and Chee Pin Tan$^{1,3}$
\thanks{$^{1}$The authors are with the School of Engineering, Monash University Malaysia, Bandar Sunway, 47500 Selangor, Malaysia (E-mail: loo.junnyong/ding.zeyang/surya.nurzaman/tan.chee.pin/@monash.edu).}%
\thanks{$^{2}$The authors are with the School of Information Technology, Monash University Malaysia, Bandar Sunway, 47500 Selangor, Malaysia (E-mail: vishnu.monn@monash.edu).
$^{3}$Corresponding author.}
\thanks{The work of Junn Yong Loo is supported by Monash University under the SIT Collaborative Research Seed Grants 2024 I-M010-SED-000242.}
}

\maketitle
 
\begin{abstract}
Most works on joint state and unknown input (UI) estimation require the assumption that the UIs are linear; this is potentially restrictive as it does not hold in many intelligent autonomous systems. To overcome this restriction and circumvent the need to linearize the system, we propose a derivative-free Unknown Input Sigma-point Kalman Filter (SPKF-nUI) where the SPKF is interconnected with a general nonlinear UI estimator that can be implemented via nonlinear optimization and data-driven approaches. The nonlinear UI estimator uses the posterior state estimate which is less susceptible to state prediction error. In addition, we introduce a joint sigma-point transformation scheme to incorporate both the state and UI uncertainties in the estimation of SPKF-nUI. An in-depth stochastic stability analysis proves that the proposed SPKF-nUI yields exponentially converging estimation error bounds under reasonable assumptions. Finally, two case studies are carried out on a simulation-based rigid robot and a physical soft robot, i.e., robots made of soft materials with complex dynamics to validate effectiveness of the proposed filter on nonlinear dynamic systems. Our results demonstrate that the proposed SPKF-nUI achieves the lowest state and UI estimation errors when compared to the existing nonlinear state-UI filters.
\end{abstract}



\definecolor{limegreen}{rgb}{0.2, 0.8, 0.2}
\definecolor{forestgreen}{rgb}{0.13, 0.55, 0.13}
\definecolor{greenhtml}{rgb}{0.0, 0.5, 0.0}

\section{Introduction}
\IEEEPARstart{R}{apid} evolution on industrial instrumentation, computing and communications in recent years have facilitated a growing thirst for intelligent autonomous systems that perceives both the representations of the physical system and its surrounding environment.
These perceptions rely heavily on information of the system's internal states and the external excitations (which is represented by {\em unknown inputs} (UIs)).
In this context, state and UI estimations are crucial in realizing accurate perceptions, which are critical for appropriate closed-loop decisions and actions in complex autonomous system.
Joint state and UI estimation has been well-established for linear continuous-time systems, 
but not for nonlinear discrete-time systems. 
In fact, most modern systems are inherently complex and nonlinear; and most estimation schemes are practically implemented in discrete time.

Initially, Kitanidis \cite{kitanidis} developed a minimum-variance unbiased Kalman filter (KF-MVU) based on the assumption that no information about the unknown input is available, in decoupling effect of the non-estimated UIs from the state estimation.
Gillijns and Moor \cite{gillijns} developed a KF-MVU that considered joint MVU unknown input and state estimation, with the UI estimation obtained via weighted least-squares.
Zhou et al. \cite{Zhisong} extended the KF-MVU to simultaneously estimate the states, UI (steering angle) and parameters of a preceding target vehicle.
Yu et al. \cite{dongdong} developed a distributed KF for cyber-physical systems with UIs and delayed measurements, where the UIs are modelled as random variables with non-informative prior distribution.
For nonlinear systems, the Extended Kalman Filter (EKF) locally linearizes the nonlinear models with respect to (w.r.t.) the estimated states before applying the KF updates.
An EKF with recursive least-squares UI estimation (EKF-UI) was first introduced in \cite{yang,huang} for structural control applications. 
Ghahremani and Kamwa \cite{Ghahremani} applied this EKF-UI to estimate the states and UI (exciter output voltage) of a power system.
Recently, Joseph et al. \cite{joseph} applied an EKF extension of the KF-MVU in \cite{kitanidis} similarly to synchronous power system, albeit without estimating the UIs.
Wei et al. \cite{Wenpeng} implemented the EKF-UI to estimate the states and UI (clutch torque) of a vehicle system.

Despite the success in various applications, the EKF generally performs poorly on highly nonlinear systems due to the large linearization errors \cite{julier}.
As a derivative-free alternative to the EKF, the Sigma-point Kalman Filters (SPKFs), i.e., Cubature Kalman Filter (CKF) \cite{arasaratnam} and Unscented Kalman Filter (UKF) \cite{julier} estimate the model-transformed mean and covariance up to higher order terms in the Taylor series expansion.
To incorporate UI estimation in the SPKFs, Anagnostou and Pal \cite{anagnostou} applied the conventional UKF with a two-stage covariance prediction alongside the least-squares UI estimation for power system application. Zheng et al. \cite{zhengzhao1,zhengzhao2} developed two UKF extensions of the KF-MVU in \cite{gillijns}, where the least-squares UI estimation is performed on top of the statistical linearization provided by the UKF.
Recently, Xue et al. \cite{Zhongjin} developed a robust M-estimation-based UKF to estimate the states and UIs (steering torque) of a vehicle system, with the UI estimation performed via iteratively reweighted least-squares.
Jiang et al. \cite{jiangkai} applied an extended-state UKF for motor-transmission systems, where the UIs are regarded as the extended states.
Kim et al. \cite{Jaehoon} applied an adaptive extended-state UKF based on selective scaling for overhead cranes, and similarly treating the UIs as extended states.

Despite the efforts in establishing joint state-UI estimation, UI estimations of these existing approaches are largely based on linear least-squares which relies on having linear models.
On one hand, approaches based on the EKF \cite{yang,huang,Ghahremani,joseph,Wenpeng} achieved this via first-order local linearization, which introduces large linearization errors and leads to poor filter estimation in the case of highly nonlinear systems. 
Also, these approaches did not explicitly take into account the uncertainty of UI estimation.
On the other hand, approaches based on the SPKF \cite{zhengzhao1,zhengzhao2,anagnostou,Zhongjin} assume that the UI is linearly separable in the nonlinear system model and thus have limited applicability.
Moreover, these approaches are more computationally demanding due to having an additional state prediction and covariance update after the UI estimation.
In addition, the UI estimation in most of the existing approaches \cite{yang,huang,Ghahremani,joseph,Wenpeng,zhengzhao1,zhengzhao2,anagnostou,Zhongjin,jiangkai,Jaehoon} used the prior state estimate which is susceptible to state prediction error, thus compromising the quality of the UI estimation.
Apart from that, state-UI filters \cite{yang,huang,Ghahremani,jiangkai,Jaehoon} are applicable only to a restricted class of systems in which the UI enters the measurement model.
More importantly, the previous stability analyses on KF \cite{rhudy}, Kalman-Consensus Filter \cite{wei}, EKF \cite{reif2}, UKF \cite{xiong} and CKF \cite{xu} did not consider UI estimation.
To the best of our knowledge, general nonlinear UI estimation in Kalman filtering is still relatively unexplored.

\begin{figure}[t]
\centering
\includegraphics[width=1\columnwidth]{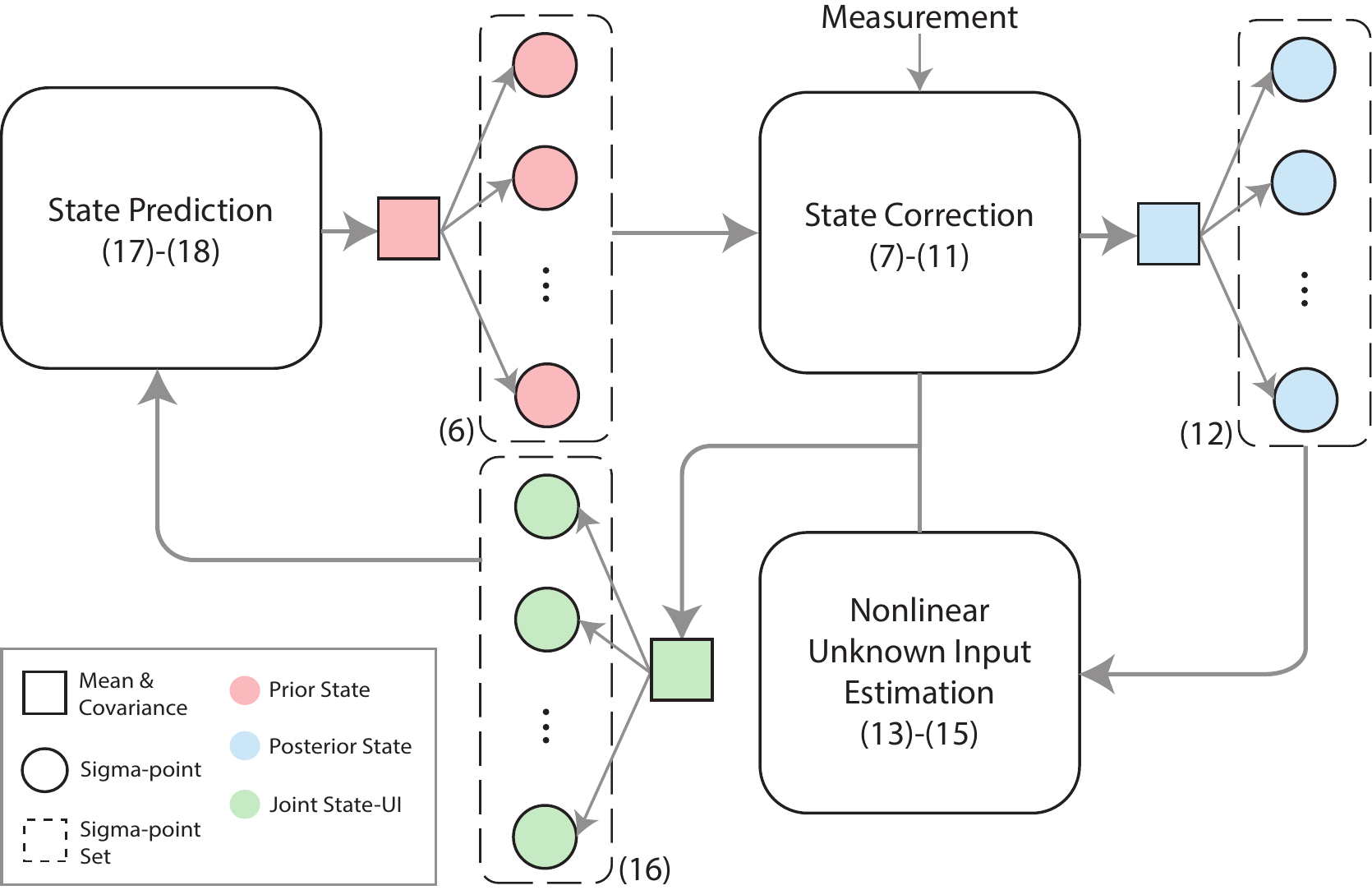}
\caption{\textbf{Illustration of the proposed SPKF-nUI.} A block diagram is used to illustrate and summarize the proposed SPKF-nUI filter (\ref{eq:init})-(\ref{eq:Pcovariance_predict}).}
\label{fig:block_diagram}
\end{figure}

Motivated by the discussion above, in this paper, we present a general derivative-free Sigma-point Kalman filter with nonlinear UI estimation (SPKF-nUI) to overcome the constraints in the existing works. 
The contributions of this paper are:
\begin{enumerate}
\item We develop a novel SPKF-nUI filtering scheme that involves an additional phase of UI estimation, which employs commonly available optimization or data-driven approaches in constructing a nonlinear UI estimator that predicts the UI from the more accurate posterior state estimate.
Existing approaches require the UI to be linearly separable \cite{zhengzhao1,zhengzhao2,anagnostou,Zhongjin}, or the UI to enter the measurement model \cite{yang,huang,Ghahremani,jiangkai,Jaehoon}; our proposed approach dispenses with these requirements.
\item We modify the conventional SPKF to include a sigma-point transformation scheme that account for the joint uncertainty in state and UI estimations, which is not considered in \cite{yang,huang,Ghahremani,joseph,Wenpeng}. The proposed scheme applies sigma-point transformation on top of the nonlinear UI estimator to generate a set of UI sigma-point estimations, which are then concatenated with the existing state sigma-points in computing the joint state-UI covariance. This allows the incorporation of the UI uncertainties into the state prediction phase of SPKF-nUI, thus enhancing its robustness against the UI estimation error.
\item Lastly, we conduct an in-depth stochastic stability analysis of the proposed SPKF-nUI on differentiable nonlinear models. Our analysis shows that the state and UI estimation errors of the SPKF-nUI are exponentially bounded in mean-square, amidst model and measurement noises with bounded covariances. 
Also, we justify the advantage of the SPKF over the EKF based on implications of the remainders in the Taylor series expansions.
\end{enumerate}

To demonstrate the effectiveness of the proposed SPKF-nUI, it is verified on two case studies with unique and complex dynamics. The first case study is a rigid robot simulation with analytical state-space model, and a nonlinear UI optimization is considered in this example.
The second case study is a physical soft robot, i.e., highly complex robot made of soft compliance materials \cite{junn_soro}. Considering the difficulty in develop an analytical model for soft robots, a class of deep learning architecture, recurrent neural network (RNN) is used to implicitly identify the nonlinear system and UI models. An example of combining deep learning and SPKF has been demonstrated in \cite{liyanjun}, albeit without UI estimation.
The case study results show that our proposed SPKF-nUI outperforms the conventional filters in state and UI estimations.

The rest of the paper is organized as follows. Section \ref{sect:SPKF-nUI} outlines the proposed SPKF-nUI. Section \ref{sect:StabilityAnalysis} provides an in-depth stochastic stability analysis on the proposed filter. The case studies demonstrating the SPKF-nUI are detailed in Section \ref{sect:IllustrativeExample}. The results are presented and discussed in Section \ref{sect:ResultsandDiscussions}. Finally, Section \ref{sect:Conclusion} concludes the paper.

\section{Unscented Kalman Filter with Nonlinear Unknown Input Estimation} \label{sect:SPKF-nUI}
\newtheorem{definition}{Definition}
Consider the following stochastic nonlinear discrete system:
\begin{align}
\begin{split} \label{eq:ss_disc}
x_{t+1} &= f(x_{t},u_{t}) + w_{t} \\
y_{t} &= h(x_{t}) + v_{t},
\end{split}
\end{align}
where $x \in {\mathbb{R}}^{n}$ is the state, $y \in {\mathbb{R}}^{m}$ is the measured output, $u \in {\mathbb{R}}^{d}$ is the UI, and $t \in \mathbb{Z}_{\geq 0}$ is the time sample. 
Here, $w \in {\mathbb{R}}^{n}$ and $v \in {\mathbb{R}}^{m}$ represent process and measurement noise, respectively.
Notice that in (\ref{eq:ss_disc}), the UI is not linearly separable from the state model $f: {\mathbb{R}}^{n+d} \rightarrow {\mathbb{R}}^{n}$; thus existing nonlinear state-UI filters \cite{zhengzhao1,zhengzhao2,anagnostou,Zhongjin} are not applicable here. 
Also, unlike existing filters \cite{yang,huang,Ghahremani,jiangkai,Jaehoon}, here the UI is not required to enter the system (\ref{eq:ss_disc}) via measurement model $h: {\mathbb{R}}^{n} \rightarrow {\mathbb{R}}^{m}$.

On top of that, consider the nonlinear UI optimization:
\begin{align} \label{eq:ui_opt}
u_{t} = \underset{u_{t}}{\argmin} \, \| \Phi(x_{t},u_{t}) \|^2 + \varepsilon_{t},
\end{align}
where $\| \cdot \|$ denotes the $\mathcal{L}^2$ norm, and $\Phi: {\mathbb{R}}^{n+d} \rightarrow {\mathbb{R}^l}$ is a nonlinear residual function. In this context, $\varepsilon_{t} \in {\mathbb{R}}^{d}$ represents errors arising from the modeling assumption on $\Phi$ or a non-convex optimization.
The nonlinear residual function $\Phi$ can generally be formulated
by imposing a zero-order hold $x_{t+1} = x_{t}$ on the states of (\ref{eq:ss_disc}) to yield
$\Phi(x_{t},u_{t}) = x_{t} - f(x_{t},u_{t})$.
Alternatively, when training data are available, a nonlinear UI model $\phi: {\mathbb{R}}^{n} \rightarrow {\mathbb{R}}^{d}$ can be implicitly identified using data-driven approaches as follows:
\begin{align} \label{eq:ui_model}
u_{t} = \phi(x_{t}) + \varepsilon_{t}.
\end{align}
where $\varepsilon_{t}$ represents modelling error of the data-driven UI model.
The formulation in (\ref{eq:ui_opt}) or (\ref{eq:ui_model}) is then used to estimate $u_{t}$ (see (\ref{eq:uiopt_posterior})).
In this paper, we demonstrate two case studies where the UIs are estimated, respectively, via solving nonlinear least-squares optimization (\ref{eq:ui_opt}) (Section \ref{sect:IllustrativeExample}-A) and directly from data-driven UI model (\ref{eq:ui_model}) (Section \ref{sect:IllustrativeExample}-B) parameterized by deep neural networks.


Now we present the algorithm of the proposed Unknown-Input (nonlinear) Sigma-point Kalman Filter (SPKF-nUI). 
Our proposed SPKF-nUI consists of the following steps. In the following, $\mathrm{Q}_{t}$, $\mathrm{R}_{t}$, $\mathrm{E}_{t}$ denote the known (available) filter parameters, in contrast to the actual covariances of $w_{t}$, $v_{t}$, $\varepsilon_{t}$ (see (\ref{eq:noise_cov})), which could be unknown.

\noindent 
\textit{\underline{Step 1 - Initialization}}:
SPKF-nUI is initialized with
\begin{align} \label{eq:init}
&\hat{x}^-_{0} = \mean[x_0], \quad 
\hat{\mathrm{P}}^{{xx}^-}_{0} = \mean[\tilde{x}^-_{0} \tilde{x}^{-^T}_{0}],
\end{align}
where $\tilde{x}^-_{0} := x_0 - \hat{x}^-_{0}$.

\noindent
\textit{\underline{Step 2 - State Correction}}:
\begin{align}
&{x}^-_{i,t} = \mathrm{sgms}\,(\hat{x}^-_{t},\hat{\mathrm{P}}^{{xx}^-}_{t}), \label{eq:sgms_prior}\\
&{y}_{i,t} = h({x}^-_{i,t}), \quad
\hat{y}_{t} = \sum_{i=0}^{2n} W_{i} \, {y}_{i,t}, \label{eq:prior_meas} \\
&\mathrm{\hat{P}}^{{xy}}_{t} = \sum_{i=0}^{2n}  W_{i} ({x}^-_{i,t} - \hat{x}_{t}^-)({y}_{i,t} - \hat{y}_{t})^T, \label{eq:Pxycovariance_prior} \\
&\mathrm{\hat{P}}^{{yy}}_{t} = \sum_{i=0}^{2n}  W_{i} ({y}_{i,t} - \hat{y}_{t})({y}_{i,t} - \hat{y}_{t})^T + \mathrm{R}_{t}, \label{eq:Pyycovariance_prior}\\
&\hat{x}_{t} = \hat{x}_{t}^- + \mathrm{K}_{t} (y_{t} - \hat{y}_{t}), \quad
\mathrm{K}_{t} = \mathrm{\hat{P}}^{xy}_{t} \, \mathrm{\hat{P}}^{{yy}^{-1}}_{t}, \label{eq:Kalman_gain} \\
&\hat{\mathrm{P}}^{xx}_{t} = \hat{\mathrm{P}}^{{xx}^-}_{t} - \mathrm{K}_{t} \mathrm{\hat{P}}^{{yy}}_{t} \mathrm{K}^{T}_{t}. \label{eq:Pcovariance_posterior}
\end{align}
where the sigma-point generation $\mathrm{sgms}\,(x,\mathrm{P})$ is defined as
\begin{align} \label{eq:sigma_points}
\begin{array}{ll}
x_{i} = x, & i = 0, \\
x_{i} = x + \sqrt{n+a}\;(\sqrt{\mathrm{P}}\:)_i, & i = 1,\dots,n, \\
x_{i} = x - \sqrt{n+a}\;(\sqrt{\mathrm{P}}\:)_{i-n}, & i = n+1,\dots,2n,
\end{array}
\end{align}
where $a \in \mathbb{R}_{\geq 0}$ is a tuning parameter. $({\mathrm{A}})_i$ denotes the $i^{th}$ column of matrix $\mathrm{A}$, and $\sqrt{\mathrm{A}}$ is the square root decomposition of $\mathrm{A}$ such that $\mathrm{A} = \sqrt{\mathrm{A}}\sqrt{\mathrm{A}}^T$. 

\noindent 
\textit{\underline{Step 3 - Nonlinear UI Estimation}}:
\begin{align}
&{x}^{+}_{i,t} = \mathrm{sgms}\,(\hat{x}_{t},\mathrm{\hat{P}}^{xx}_{t}), \label{eq:sgms_posterior}\\
\begin{split} \label{eq:uiopt_posterior}
&{u}^{+}_{i,t} = \underset{u_{t}}{\argmin} \| \Phi({x}^{+}_{i,t},u_{t}) \|^2 \;\; \text{or} \;\; \phi({x}^{+}_{i,t}), \quad
\hat{u}_{t} = \sum_{i=0}^{2n} W_{i} \, {u}^{+}_{i,t}, \\
\end{split} \\
&\mathrm{\hat{P}}^{xu}_{t} = \mathrm{\hat{P}}^{{ux}^T}_{t} = \sum_{i=0}^{2n}  W_{i} ({x}^{+}_{i,t} - \hat{x}_{t})({u}^{+}_{i,t} - \hat{u}_{t})^T, \label{eq:Pxucovariance_posterior} \\
&\mathrm{\hat{P}}^{uu}_{t} = \sum_{i=0}^{2n}  W_{i} ({u}^{+}_{i,t} - \hat{u}_{t})({u}^{+}_{i,t} - \hat{u}_{t})^T + \mathrm{E}_{t}. \label{eq:Puucovariance_posterior}
\end{align}
Here, we apply sigma-point transformations to the UI estimator in (\ref{eq:uiopt_posterior}), followed by the covariance estimations in (\ref{eq:Pxucovariance_posterior})-(\ref{eq:Puucovariance_posterior}).
Notice that the UI estimator (\ref{eq:uiopt_posterior}) uses the posterior sigma-points ${x}^{+}_{i,t}$ generated via (\ref{eq:sgms_posterior}), instead of the less accurate prior state used in existing approaches \cite{yang,huang,Ghahremani,joseph,Wenpeng,zhengzhao1,zhengzhao2,anagnostou,Zhongjin}.

\noindent 
\textit{\underline{Step 4 - State Prediction}}:
\begin{align}
\begin{split} \label{eq:sgms_joint}
&\begin{bmatrix} {x}^{+}_{i,t} \\ {u}^{+}_{i,t} \end{bmatrix} =  \mathrm{sgms} \bigg( \begin{bmatrix} \hat{x}_{t} \\ \hat{u}_{t} \end{bmatrix}, \mathrm{\hat{P}}^{xxuu}_{t} 
\bigg), \quad
\mathrm{\hat{P}}^{xxuu}_{t}
= \begin{bmatrix} \mathrm{\hat{P}}^{xx}_{t} & \mathrm{\hat{P}}^{xu}_{t} \\ \mathrm{\hat{P}}^{ux}_{t} & \mathrm{\hat{P}}^{uu}_{t} \end{bmatrix}
\end{split} \\
&{x}^-_{i,t+1} = f({x}^{+}_{i,t},{u}^{+}_{i,t}), \quad 
\hat{x}_{t+1}^- = \sum_{i=0}^{2n} W_{i} \, {x}^-_{i,t+1}, \label{eq:prior_state} \\
&\hat{\mathrm{P}}^{{xx}^-}_{t+1} = \sum_{i=0}^{2n}  W_{i} ({x}^-_{i,t+1} - \hat{x}_{t+1}^-)({x}^-_{i,t+1} - \hat{x}_{t+1}^-)^T + \mathrm{Q}_{t}. \label{eq:Pcovariance_predict}
\end{align}
Here, a new set of sigma-points $\{{x}^{+}_{i,t}, {u}^{+}_{i,t}\}_{i=0}^{2n}$ is generated in (\ref{eq:sgms_joint}) using the joint state-UI covariance $\mathrm{\hat{P}}^{xxuu}_{t}$ computed via (\ref{eq:uiopt_posterior})-(\ref{eq:Puucovariance_posterior}). As such, this new sigma-point ensemble encapsulates the joint state-UI uncertainties that can be incorporated into the state prediction (\ref{eq:prior_state})-(\ref{eq:Pcovariance_predict}). 


\noindent 
\textit{\underline{Step 5 - Repeat}}: Set $t = t+1$ and repeat steps 2 to 4. \\

To sum up, our proposed SPKF-nUI algorithm (\ref{eq:init})-(\ref{eq:Pcovariance_predict}) has several advantages over existing approaches. First, it allows the UI $u_t$ to enter the nonlinear state model $f$, which is not possible in \cite{zhengzhao1,zhengzhao2,anagnostou,Zhongjin} due to the UI being linearly separable. Also, an additional covariance parameter $\mathrm{E}_{t}$ is incorporated in (\ref{eq:Puucovariance_posterior}) to account for the UI noise $\varepsilon_t$ (which also enters the state model, and could degrade the model prediction if neglected). In addition, the state-UI covariance (\ref{eq:Pxucovariance_posterior}) incorporates joint state-UI uncertainty (not considered in \cite{yang,huang,Ghahremani,joseph,Wenpeng}) to achieve more accurate nonlinear filtering and predictive uncertainty characterization. Furthermore, the additional UI estimation steps allows the SPKF-nUI to perform state prediction in two stages (\ref{eq:sgms_posterior})-(\ref{eq:uiopt_posterior}) (Step 3) and (\ref{eq:sgms_joint})-(\ref{eq:prior_state})  (Step 4); this reduces the sigma-point approximation error by circumventing a compounded sigma transformation, resulting from a model composition of the state model $f(x_{t}, u_{t})$ and the UI estimation in (\ref{eq:ui_opt}) or (\ref{eq:ui_model}), where model nonlinearity is intensified.
A block diagram is shown in Fig. \ref{fig:block_diagram} to summarize the proposed filter.


\section{Stochastic Stability Analysis} \label{sect:StabilityAnalysis}
In this section, we present a stability analysis of the proposed SPKF-nUI on twice differentiable (everywhere) nonlinear systems and UI model functions. First, we acquire the update equations for the state estimation error $\tilde{x}_{t}$ in Section \ref{ssect:StateErrorPropagation}.
Then, we simplify the update equations for the Kalman gain $\mathrm{K}_{t}$ and the posterior state covariance $\hat{\mathrm{P}}^{xx}_{t}$ in Section \ref{ssect:CovarianceEstimateUpdate}
Lastly, we prove the exponential error boundedness of SPKF-nUI via a Lyapunov-based stability analysis in Section \ref{ssect:MainStabilityResults}.

\subsection{State Error Propagation} \label{ssect:StateErrorPropagation}
In this subsection, we formulate an update equation for the posterior state error $\tilde{x}_{t}$.
Here, we follow the notation used in \cite{folland}, where $\alpha = \left( \alpha_1,\dots,\alpha_n \right)$ and $\alpha_j \in \mathbb{Z}_{\geq 0}$, and we have
\begin{align} 
\begin{split} \label{eq:notation_folland}
|\alpha| = \sum_{i=1}^n \alpha_i, &\qquad
\alpha! = \prod_{i=1}^n \alpha_i!, \\
x^\alpha = \prod_{i=1}^n x_i^{\alpha_i}, &\qquad
\partial^\alpha = \prod_{i=1}^n \big( \frac{\partial}{\partial x_i} \big)^{\alpha_i}.
\end{split} 
\end{align}
\newtheorem{lemma}{Lemma}
\begin{lemma} [Multivariate Taylor Series Expansion] \cite{folland} \label{lemma1} \\
Suppose $f := \begin{bmatrix} f_1,\dots,f_m \end{bmatrix}^T :\mathbb{R}^n \rightarrow \mathbb{R}^m$ and each $f_i :\mathbb{R}^n \rightarrow \mathbb{R}$ is of class $C^{k}$, i.e., $k$-times differentiable on an open convex set $S$. If $\hat{x} \in S$ and $x = \hat{x} + \tilde{x} \in S$ and, then
\begin{subequations}
\begin{align}
f(x) = \sum_{|\alpha| \leq \, {k-1}} \frac{\partial^\alpha f(\hat{x})}{\alpha!} \, \tilde{x}^{\alpha} + R_{f,\hat{x}}^{k}(\tilde{x})
\end{align}
with $k \in \mathbb{Z}_{\geq 0}$, and the Taylor remainder is given by
\begin{align}
R_{f,\hat{x}}^{k}(\tilde{x}) = \sum_{|\alpha| = k} \frac{\partial^\alpha f(\hat{x}+c\tilde{x})}{\alpha!} \, \tilde{x}^{\alpha}
\end{align}
for some $c \in \left( 0,1 \right)$.
\end{subequations}
\end{lemma}

Denote the prior state error, posterior state error, UI error and innovation, respectively as
\begin{align} 
\begin{split} 
\tilde{x}_{t}^- := x_{t} - \hat{x}_{t}^-, &\qquad
\tilde{x}_{t} := x_{t} - \hat{x}_{t}, \\
\tilde{u}_{t} := u_{t} - \hat{u}_{t}, &\qquad
\tilde{y}_{t} := y_{t} - \hat{y}_{t}.
\end{split}
\end{align}
and denote the state model's argument variables as ${X}_t = \begin{bmatrix} {x}_{t}^T & {u}_{t}^T \end{bmatrix}^T$. 
Assume that the state model $f$ in (\ref{eq:ss_disc}) is twice differentiable everywhere. According to Lemma \ref{lemma1}, the Taylor series expansion of $f$ about $\hat{X}_t = \begin{bmatrix} \hat{x}_{t}^T & \hat{u}_{t}^T \end{bmatrix}^T$ at $k=1$ gives
\begin{align} 
\begin{split} \label{eq:Taylor_AbuseOfNotation}
x_{t+1} &= \sum_{|\alpha| \leq 1} \frac{\partial^\alpha f(\hat{X}_{t})}{\alpha!} \, \tilde{X}_{t}^{\alpha} + R_{f,\hat{X}_{t}}^2(\tilde{X}_{t}) + w_{t},
\end{split}
\end{align}
where $\tilde{X}_{t} = {X}_t - \hat{X}_t$.
Based on the notations in (\ref{eq:notation_folland}), the first term of (\ref{eq:Taylor_AbuseOfNotation}) can be written as
\begin{align*} 
\begin{split}
\sum_{|\alpha| \leq 1} \frac{\partial^\alpha f(\hat{X}_{t})}{\alpha!} \, \tilde{X}_{t}^{\alpha} = f(\hat{X}_{t}) + \mathrm{F}_{t} \tilde{x}_{t} + \mathrm{G}_{t} \tilde{u}_{t}
\end{split}
\end{align*}
with
$\mathrm{F}_t = \frac{\partial f}{\partial x_{t}} \Big|_{(\hat{x}_{t},\hat{u}_{t})}$, 
$\mathrm{G}_t = \frac{\partial f}{\partial u_{t}} \Big|_{(\hat{x}_{t},\hat{u}_{t})}$,
and (\ref{eq:Taylor_AbuseOfNotation}) becomes
\begin{align} 
\begin{split} \label{eq:xt_Taylor}
x_{t+1} &= f(\hat{X}_{t}) + \mathrm{F}_{t} \tilde{x}_{t} + \mathrm{G}_{t} \tilde{u}_{t} + R_{f,\hat{X}_{t}}^2(\tilde{X}_{t}) + w_{t}.
\end{split}
\end{align}
Subtract $\hat{x}_{t+1}$ of (\ref{eq:Kalman_gain}) from (\ref{eq:xt_Taylor}) gives the posterior state error
\begin{align}
\begin{split} \label{eq:x_error_expand}
\tilde{x}_{t+1} &= x_{t+1} - (\hat{x}_{t+1}^- + \mathrm{K}_{t+1}\tilde{y}_{t+1})
= \tilde{x}_{t+1}^- - \mathrm{K}_{t+1}\tilde{y}_{t+1}.
\end{split}
\end{align}
Expanding the prior state estimate of (\ref{eq:prior_state}) gives
\begin{align} \label{eq:Taylor_AbuseOfNotation_prior}
\hat{x}_{t+1}^- =& \frac{a}{n + a} \, f({X}_{0,t}) + \frac{1}{2(n + a)} \sum_{i=1}^{2n} f({X}_{i,t}),
\end{align}
where ${X}_{i,t} = \begin{bmatrix} {{x}^{+T}_{i,t}} & {u}^{+T}_{i,t} \end{bmatrix}^T$. 
A Taylor series expansion of $f({X}_{i,t})$ about $\hat{X}_{t}$ then yields
\begin{align} \label{eq:xt-_Taylor}
\hat{x}_{t+1}^- = f(\hat{X}_{t}) + \frac{1}{2(n + a)} \sum_{i=1}^{2n} R_{f,\hat{X}_{t}}^2(\tilde{{X}}_{i,t}).
\end{align}
where $\tilde{{X}}_{i,t} = {X}_{i,t} - \hat{X}_{t}$. In particular, the first-order term vanishes as the set of sigma-points cancel out each other in the summation due to their symmetry defined in (\ref{eq:sigma_points}).
Subtracting (\ref{eq:xt-_Taylor}) from (\ref{eq:xt_Taylor}) gives the prior state error
\begin{align}
\begin{split} \label{eq:prior_x_error}
\tilde{x}_{t+1}^- &= \mathrm{F}_{t} \tilde{x}_{t} + \mathrm{G}_{t} \tilde{u}_{t} + {r_f}(\tilde{x}_{t},\tilde{u}_{t}) + w_{t},
\end{split}
\end{align}
and the state remainder error 
\begin{align}
\begin{split} \label{eq:residual_phi}
{r_f}(\tilde{x}_{t},\tilde{u}_{t}) = R_{f,\hat{X}_{t}}^2(\tilde{X}_{t}) - \frac{1}{2(n + a)} \sum_{i=1}^{2n} R_{f,\hat{X}_{t}}^2(\tilde{{X}}_{i,t}).
\end{split}
\end{align}
in the second-order Taylor remainder terms of (\ref{eq:xt_Taylor}) and (\ref{eq:xt-_Taylor}).

Assuming that the residual function $\Phi$ in (\ref{eq:ui_opt}) and the model $\phi$ in (\ref{eq:ui_model}) are twice differentiable everywhere, expanding and subtracting $\hat{u}_{t}$ of (\ref{eq:uiopt_posterior}) from ${u}_{t}$ of (\ref{eq:ui_opt}) or (\ref{eq:ui_model}) gives the UI error
\begin{align} \label{eq:post_u_error}
\begin{split}
\tilde{u}_{t} &= \mathrm{M}_{t} \tilde{x}_{t} + {r_\phi}(\tilde{x}_{t}) + \varepsilon_{t}
\end{split}
\end{align}
where
\begin{align} \label{eq:jacobian_u}
\begin{split}
&\mathrm{M}_{t} 
= \bigg(\frac{\partial \Phi}{\partial u_{t}}\bigg)^{\!\!\dagger} \frac{\partial \Phi}{\partial x_{t}} \Big|_{\hat{x}_{t}} \quad \text{or} \quad \mathrm{M}_{t} = \frac{\partial \phi}{\partial x_{t}} \Big|_{\hat{x}_{t}}
\end{split}
\end{align}
of which the left-hand case is obtained via applying the Gauss–Newton method \cite{chong} on the UI optimization (\ref{eq:ui_opt}), and the right-hand case corresponds to having a data-driven UI model (\ref{eq:ui_model}). Here, $\mathrm{A}^{\dagger}$ denotes the Moore–Penrose inverse of matrix $\mathrm{A}$.
Subsequently, the UI remainder error is given by
\begin{align}
\begin{split} \label{eq:residual_tau}
{r_\phi}(\tilde{x}_{t}) = R_{\phi,\hat{x}_{t}}^2(\tilde{x}_{t}) - \frac{1}{2(n + a)} \sum_{i=1}^{2n} R_{\phi,\hat{x}_{t}}^2(\tilde{x}_{i,t}).
\end{split}
\end{align}
By substituting (\ref{eq:post_u_error}) into (\ref{eq:prior_x_error}), the prior state error becomes
\begin{align}
\begin{split} \label{eq:prior_x_only_error}
\tilde{x}_{t+1}^- &= \mathrm{J}_{t} \tilde{x}_{t} + {r_f}(\tilde{x}_{t},\tilde{u}_{t}) + \mathrm{G}_{t} {r_\phi}(\tilde{x}_{t}) + w_{t} + \mathrm{G}_{t} \varepsilon_{t},
\end{split}
\end{align}
where $\mathrm{J}_{t} = \mathrm{F}_{t} + \mathrm{G}_{t} \mathrm{M}_{t}$.

Assuming that the measurement model $h$ in (\ref{eq:ss_disc}) is twice differentiable everywhere, expand and subtract $\hat{y}_{t}$ of (\ref{eq:prior_meas}) from ${y}_{t}$ of (\ref{eq:ss_disc}) gives the innovation
\begin{align}
\begin{split} \label{eq:prior_y_error}
\tilde{y}_{t} &= \mathrm{H}_{t} \tilde{x}^-_{t} + {r_h}(\tilde{x}_{t}^-) + v_{t}, \quad
\mathrm{H}_{t} = \frac{\partial h}{\partial x_{t}} \Big|_{\hat{x}^-_{t}}
\end{split}
\end{align}
and the measurement remainder error
\begin{align}
\begin{split} \label{eq:residual_varphi}
{r_h}(\tilde{x}^-_{t}) = R_{h,\hat{x}_{t}}^2(\tilde{x}^-_{t}) - \frac{1}{2(n + a)} \sum_{i=1}^{2n} R_{h,\hat{x}_{t}^-}^2(\tilde{x}^-_{i,t}).
\end{split}
\end{align}
Finally, by substituting (\ref{eq:prior_x_only_error}), (\ref{eq:prior_y_error}) into (\ref{eq:x_error_expand}), we obtain the update equation of posterior state error as follows:
\begin{subequations} \label{eq:x_posterior_errors}
\begin{align}
\begin{split} \label{eq:x_posterior_error}
&\tilde{x}_{t+1} 
= \mathrm{L}_{t+1} \mathrm{J}_{t} \tilde{x}_{t} + r_{t} + s_{t},
\end{split}
\\
\begin{split} \label{eq:r_t}
&r_{t} = \mathrm{L}_{t+1} \big( {r_f}(\tilde{x}_{t},\tilde{u}_{t}) + \mathrm{G}_{t} {r_\phi}(\tilde{x}_{t}) \big) 
- \mathrm{K}_{t+1} {r_h}(\tilde{x}_{t+1}^-),
\end{split} \\
\begin{split} \label{eq:s_t}
&s_{t} = \mathrm{L}_{t+1} \left( w_{t} + \mathrm{G}_{t} \varepsilon_{t} \right) - \mathrm{K}_{t+1}v_{t+1},
\end{split}
\end{align}
\end{subequations}
where $\mathrm{L}_{t} = \mathrm{I} - \mathrm{K}_{t}\mathrm{H}_{t}$. Here, $r_{t}$ and $s_{t}$ encapsulate the nonlinear terms and the noise terms, respectively.

\subsection{Kalman Gain and Covariance Updates} \label{ssect:CovarianceEstimateUpdate}

In this subsection, we simplify the update equations of the Kalman gain $\mathrm{K}_{t}$ and the posterior state covariance $\hat{\mathrm{P}}^{xx}_{t}$.
The Taylor series expansion of the prior (state) sigma-point deviations ${x}^-_{i,t+1} - \hat{x}_{t+1}^-$ of $\hat{\mathrm{P}}^{{xx}^-}_{t+1}$ in (\ref{eq:Pcovariance_predict}) gives
\begin{subequations} \label{eq:sigma-point deviation}
\begin{align}
\begin{split} \label{eq:sigma-point deviation_a}
{x}^-_{0,t+1} - \hat{x}_{t+1}^- =& -\frac{1}{2(n + a)} \sum_{i=1}^{2n} R_{f,\hat{X}_{t}}^2(\tilde{{X}}_{i,t}),
\end{split} \\
\begin{split} \label{eq:sigma-point deviation_b}
{x}^-_{i,t+1} - \hat{x}_{t+1}^- =&\; \sum_{|\alpha| = 1} \frac{\partial^\alpha f(\hat{X}_{t})}{\alpha!} \, \tilde{{X}}_{i,t}^{\alpha} + R_{f,\hat{X}_{t}}^2(\tilde{{X}}_{i,t}) \\
&\;\,-\frac{1}{2(n + a)} \sum_{i=1}^{2n} R_{f,\hat{X}_{t}}^2(\tilde{{X}}_{i,t}), \\
&\qquad\qquad\qquad\qquad\quad\;\, i=1,\dots,2n.
\end{split}
\end{align}
\end{subequations}
Based on the notations in (\ref{eq:notation_folland}), the first term of (\ref{eq:sigma-point deviation_b}) can be written as
\begin{align} 
\begin{split} \label{eq:expand_prior_covariance_1}
\sum_{|\alpha| = 1} \frac{\partial^\alpha f(\hat{X}_{t})}{\alpha!} \, \tilde{{X}}_{i,t}^{\alpha} = \begin{bmatrix} \mathrm{F}_{t} & \mathrm{G}_{t} \end{bmatrix} \tilde{{X}}_{i,t},
\end{split}
\end{align}
and therefore, we have
\begin{align} 
\begin{split} \label{eq:expand_prior_covariance}
& \frac{1}{2(n + a)} \sum_{i=1}^{2n} 
\left( \sum_{|\alpha| = 1} \frac{\partial^\alpha f(\hat{X}_{t})}{\alpha!} \tilde{{X}}_{i,t}^{\alpha} \right)
\left( \sum_{|\alpha| = 1} \frac{\partial^\alpha f(\hat{X}_{t})}{\alpha!} \tilde{{X}}_{i,t}^{\alpha} \right)^T \\
&= \frac{1}{2(n + a)} \sum_{i=1}^{2n} \begin{bmatrix}\mathrm{F}_{t} & \mathrm{G}_{t} \end{bmatrix} \tilde{{X}}_{i,t}\tilde{{X}}_{i,t}^T \begin{bmatrix}\mathrm{F}_{t} & \mathrm{G}_{t} \end{bmatrix}^T \\
&= \begin{bmatrix}\mathrm{F}_{t} & \mathrm{G}_{t} \end{bmatrix} \mathrm{\hat{P}}^{xxuu}_{t} \begin{bmatrix}\mathrm{F}_{t} & \mathrm{G}_{t} \end{bmatrix}^T,
\end{split}
\end{align}
where the joint state-UI covariance $\mathrm{\hat{P}}^{xxuu}_{t}$ is defined in (\ref{eq:sgms_joint}), and the last equation in (\ref{eq:expand_prior_covariance}) is due to
$
\frac{1}{2(n + a)} \sum_{i=1}^{2n} \tilde{{X}}_{i,t}\tilde{{X}}_{i,t}^T
= \mathrm{\hat{P}}^{xxuu}_{t}
$
according to the sigma-point definition in (\ref{eq:sigma_points}).

Substituting (\ref{eq:sigma-point deviation})-(\ref{eq:expand_prior_covariance}) into (\ref{eq:Pcovariance_predict}) and neglecting the remainder error terms, we obtain
\begin{align} \label{eq:Pxx-_expand}
\mathrm{\hat{P}}^{{xx}^-}_{t+1} &= \begin{bmatrix}\mathrm{F}_{t} & \mathrm{G}_{t} \end{bmatrix}
\mathrm{\hat{P}}^{xxuu}_{t}
\begin{bmatrix}\mathrm{F}_{t} & \mathrm{G}_{t} \end{bmatrix}^T + \mathrm{Q}_{t},
\end{align}
Repeating the same procedures on (\ref{eq:Pxucovariance_posterior})-(\ref{eq:Puucovariance_posterior}) yields
$\mathrm{\hat{P}}^{xu}_{t} = \mathrm{\hat{P}}^{xx}_{t} \mathrm{M}^T_{t}$ and 
$\mathrm{\hat{P}}^{uu}_{t} = \mathrm{M}_{t} \mathrm{\hat{P}}^{xx}_{t} \mathrm{M}^T_{t} + \mathrm{E}_{t}$.
By substituting these into (\ref{eq:Pxx-_expand}), we have
\begin{align} \label{Pxx-_prior_estimate}
\begin{split}
\mathrm{\hat{P}}^{{xx}^-}_{t+1} 
&= \mathrm{J}_{t} \mathrm{\hat{P}}^{xx}_{t} \mathrm{J}_{t}^T + \mathrm{G}_{t} \mathrm{E}_{t} \mathrm{G}_{t}^T + \mathrm{Q}_{t},
\end{split}
\end{align}
Similarly, carrying out the same procedures on (\ref{eq:Pxycovariance_prior})-(\ref{eq:Pyycovariance_prior}), we have
$\mathrm{\hat{P}}_{t}^{{xy}} = \mathrm{\hat{P}}^{{xx}^-}_{t} \mathrm{H}_{t}^T$ and 
$\mathrm{\hat{P}}_{t}^{{yy}} = \mathrm{H}_{t} \mathrm{\hat{P}}^{{xx}^-}_{t} \mathrm{H}_{t}^T + \mathrm{R}_{t}$.
Substituting these into (\ref{eq:Kalman_gain})-(\ref{eq:Pcovariance_posterior}) and re-arranging yields
\begin{align}
\begin{split} \label{eq:Kalman_linear}
&\mathrm{K}_{t} = \mathrm{\hat{P}}^{{xx}^-}_{t}\mathrm{H}_{t}^T \begin{pmatrix} \mathrm{H}_{t} \mathrm{\hat{P}}^{{xx}^-}_{t} \mathrm{H}_{t}^T + \mathrm{R}_{t} \end{pmatrix}^{-1} = \hat{\mathrm{P}}^{xx}_{t} \mathrm{H}_{t}^T \mathrm{R}_{t}^{-1},
\end{split}
\\
\label{eq:P_update}
&\hat{\mathrm{P}}^{xx}_{t} = \mathrm{L}_{t} \hat{\mathrm{P}}^{{xx}^-}_{t} \mathrm{L}_{t}^T 
+ \mathrm{K}_{t} \mathrm{R}_{t} \mathrm{K}_{t}^T 
= \mathrm{L}_{t} \hat{\mathrm{P}}^{{xx}^-}_{t}.
\end{align}

\subsection{Stochastic Stability Results} \label{ssect:MainStabilityResults}
In this subsection, we present the main stochastic stability result that prove exponential boundedness of the posterior state estimation error in the proposed SPKF-nUI.
The basis of this stochastic analysis is the modified Stochastic Stability Lemma \cite{tarn} that considers time-varying parameters.
\begin{lemma} \cite{rhudy} \label{lemma3}
Assume that there are stochastic processes $\zeta_{t}$ and $V(\zeta_{t})$, and positive real numbers $\upsilon_0,\upsilon_{t},\mu_{t}$, and $\sigma_{t} \leq 1$ such that
$V(\zeta_0) \leq \upsilon_0 \| \zeta_{t} \|^2$ and $\upsilon_{t} \| \zeta_{t} \|^2 \leq V(\zeta_{t})$, and 
\begin{align} \label{eq:lemma3_condition}
\mean\big[V_{t+1}(\zeta_{t+1}) \,|\, \zeta_{t}\big] - V_{t}(\zeta_{t}) \leq - \sigma_{t} V_{t}(\zeta_{t}) + \mu_{t}
\end{align}
holds for all $t$. Then, $\zeta_{t}$ is exponentially bounded in mean-square, i.e.,
\begin{align}
\begin{split} \label{eq:lemma3_result}
\mean\left[\| \zeta_{t} \|^2\right] \leq&\; \frac{\upsilon_0}{\upsilon_{t}} \mean\left[\| \zeta_0 \|^2\right] \prod_{i=0}^{t-1} (1-\sigma_i) \\
&+ \frac{1}{\upsilon_{t}} \sum_{i=0}^{t-1} \left[ \mu_{i} \prod_{j=i+1}^{t-1} (1-\sigma_{j}) \right].
\end{split}
\end{align}
\end{lemma}

The following theorem states the main result of our analysis.
\newtheorem{theorem}{Theorem}
\begin{theorem} \label{theorem1}
Consider the nonlinear stochastic system (\ref{eq:ss_disc}) and the nonlinear UI optimization (\ref{eq:ui_opt}) or data-driven model (\ref{eq:ui_model}). For every time $t$, assume the following:

\begin{itemize}

\item
The noises $w_{t}$, $v_{t}$ and $\varepsilon_{t}$ in (\ref{eq:ss_disc})-(\ref{eq:ui_model}) are Gaussian, and mutually uncorrelated. 
Also, there exist scalars $\delta_w, \delta_v, \delta_\varepsilon \in \mathbb{R}_{>0}$ such that their covariances are bounded by
\begin{align}
\begin{split} \label{eq:noise_cov}
\mean[w_{t}w_{t}^T] \leq \delta_w \mathrm{I}_n, \quad 
\mean[v_{t}v_{t}^T] \leq \delta_v \mathrm{I}_m, \quad
\mean[\varepsilon_{t}\varepsilon_{t}^T] \leq \delta_\varepsilon \mathrm{I}_d.
\end{split}
\end{align}

\item 
There are $\overline{f}_{t}, \overline{m}_{t}, \underline{g}_{t}, \overline{g}_{t}, \underline{h}_{t}, \overline{h}_{t}, \underline{q}_{t}, \underline{e}_{t}, \underline{r}_{t}, \overline{r}_{t}, \underline{p}_{t}, \overline{p}_{t} \in \mathbb{R}_{>0}$ such that the Jacobian and Covariance matrices are bounded by
\begin{subequations} \label{eq:assump_sys}
\begin{align}
\begin{split} \label{eq:assump_FGMH}
& 
\|\mathrm{F}_{t}\| 
\leq \overline{f}_{t}, \quad
\|\mathrm{M}_{t}\| 
\leq \overline{m}_{t}, \quad
\|\mathrm{G}_{t}\| 
\leq \overline{g}_{t},
\\
&
\|\mathrm{H}_{t}\| 
\leq \overline{h}_{t}, \quad
\underline{g}_{t}^2 \mathrm{I}_n \leq 
\mathrm{G}_{t}\mathrm{G}_{t}^T, \quad 
\underline{h}_{t}^2 \mathrm{I}_m \leq 
\mathrm{H}_{t}\mathrm{H}_{t}^T, 
\end{split} 
\\
&
\underline{q}_{t} \mathrm{I}_n \leq \mathrm{Q}_{t}, \quad 
\underline{e}_{t} \mathrm{I}_d \leq \mathrm{E}_{t}, \quad
\underline{r}_{t} \mathrm{I}_m \leq \mathrm{R}_{t} \leq \overline{r}_{t} \mathrm{I}_m, \label{eq:assump_QER} 
\\
&
\underline{p}_{t} \mathrm{I}_n \leq \hat{\mathrm{P}}^{xx}_{t} \leq \overline{p}_{t} \mathrm{I}_n. \label{eq:assump_P}
\end{align}
\end{subequations}

\item 
There are $\kappa^{f}_{t}, \kappa^{\phi}_{t}, \kappa^{h}_{t} \in \mathbb{R}_{>0}$ such that the remainder errors in (\ref{eq:residual_phi}), (\ref{eq:residual_tau}) and (\ref{eq:residual_varphi}) are bounded by
\begin{subequations} \label{eq:assump_nonl}
\begin{align}
\label{eq:assump_phi}
&\| {r_f}(\tilde{x}_{t},\tilde{u}_{t}) \| 
\leq \kappa^{f}_{t} \big( \| \tilde{x}_{t} \|^2 + \| \tilde{u}_{t} \|^2 \big), \\
\label{eq:assump_tau}
&\| {r_\phi}(\tilde{x}_{t}) \|
\leq \kappa^{\phi}_{t} \| \tilde{x}_{t} \|^2, \\
\label{eq:assump_varphi}
&\| {r_h}(\tilde{x}^-_{t}) \|
\leq \kappa^{h}_{t} \| \tilde{x}^-_{t} \|^2.
\end{align}
\end{subequations}

\item 
$\mathrm{J}_{t} = \mathrm{F}_{t} + \mathrm{G}_{t} \mathrm{M}_{t}$ is full rank.

\end{itemize}
where $\| A \|$ denotes matrix norm induced by the $\mathcal{L}^2$ norm.

Then, the posterior state and UI errors $\tilde{x}_{t}$, $\tilde{u}_{t}$ of the SPKF-nUI (\ref{eq:init})-(\ref{eq:Pcovariance_predict}) are exponentially bounded in mean-square, provided that the initial error satisfies $\| \tilde{x}_{0} \| \leq \epsilon$.
\end{theorem}

\begin{proof} 
Theorem \ref{theorem1} will be proven in two stages. The first stage is Propositions \ref{proposition1}-\ref{proposition3}, where we obtain the respective bounds on the terms in the Lyapunov function. These bounds are then used in the second stage, where we prove that the state estimation error of SPKF-nUI satisfies Lemma \ref{lemma3} and thus guarantees exponential boundedness.

\newtheorem{proposition}{Proposition}
\begin{proposition} \label{proposition1}
Under assumptions in Theorem \ref{theorem1}, there exists a positive real number $\sigma_{t} \leq 1$, where $\Pi_{t} = {\hat{\mathrm{P}}_{t}}^{{xx}^{-1}}$ satisfies
\begin{align}
\begin{split} \label{eq:prop1_result}
&\mathrm{J}_{t}^T \mathrm{L}_{t+1}^T 
\Pi_{t+1} 
\mathrm{L}_{t+1} \mathrm{J}_{t} 
\leq \Pi_{t}(1 - \sigma_{t}).
\end{split}
\end{align}
\end{proposition}

\begin{proof}
Applying (\ref{eq:assump_sys}) on (\ref{eq:Kalman_linear}), we obtain the following bounds:
\begin{subequations} \label{eq:K_bounds_1}
\begin{align} 
\label{eq:KL_bound}
& \| \mathrm{K}_{t} \| \leq \frac{\overline{p}_{t}\overline{h}_{t}}{\underline{r}_{t}},
\quad
\mathrm{L}_{t} 
\leq 
\left( 1 + \frac{\overline{p}_{t}\overline{h}_{t}^2}{\underline{r}_{t}} \right) \mathrm{I}_n, \\
\label{eq:KRK_bound}
& \mathrm{K}_{t}\mathrm{R}_{t}\mathrm{K}_{t}^T 
= {\hat{\mathrm{P}}_{t}}^{xx} \mathrm{H}_{t}^T \mathrm{R}_{t}^{-1} \mathrm{H}_{t} {\hat{\mathrm{P}}_{t}}^{xx}
\geq \frac{\underline{p}_{t}^2\underline{h}_{t}^2}{\overline{r}_{t}} \mathrm{I}_n.
\end{align}
\end{subequations}
The assumption that $\mathrm{J}_{t}$ is full rank implies that $\mathrm{J}_{t}$ and $\mathrm{L}_{t}$ are invertible matrices \cite{simon}.
\begin{table*}[tb]
\normalsize
\begin{align}
\begin{split} \label{eq:P_update_reformulated_long}
\hat{\mathrm{P}}^{xx}_{t+1} 
&\;=\; \mathrm{L}_{t+1}
\hat{\mathrm{P}}^{{xx}^-}_{t+1}
\mathrm{L}_{t+1}^T + \mathrm{K}_{t+1} \mathrm{R}_{t+1} \mathrm{K}_{t+1}^T 
\;=\; \mathrm{L}_{t+1}
\big(\mathrm{J}_{t} 
\hat{\mathrm{P}}^{xx}_{t} 
\mathrm{J}_{t}^T + \mathrm{G}_{t} \mathrm{E}_{t} \mathrm{G}_{t}^T + \mathrm{Q}_{t} 
\big) 
\mathrm{L}_{t+1}^T 
+ \mathrm{K}_{t+1} \mathrm{R}_{t+1} \mathrm{K}_{t+1}^T \\
&\;=\; \mathrm{L}_{t+1} \mathrm{J}_{t}
\hat{\mathrm{P}}^{xx}_{t}
\bigg( 
\mathrm{I}_n + 
\hat{\mathrm{P}}^{{xx}^{-1}}_{t} 
\mathrm{J}_{t}^{-1} 
\, \Big(
\mathrm{G}_{t} \mathrm{E}_{t} \mathrm{G}_{t}^T 
+ \mathrm{Q}_{t} 
+ \mathrm{L}_{t+1}^{-1} \mathrm{K}_{t+1} \mathrm{R}_{t+1} \mathrm{K}_{t+1}^T \mathrm{L}_{t+1}^{-T} 
\Big) \,
\mathrm{J}_{t}^{-T}
\bigg) \,
\mathrm{J}_{t}^T \mathrm{L}_{t+1}^T 
\end{split}
\end{align}
\end{table*}
By substituting (\ref{Pxx-_prior_estimate}) into (\ref{eq:P_update}), we can expand and rearrange the posterior state covariance as shown (\ref{eq:P_update_reformulated_long}).
Applying (\ref{eq:assump_sys}) and (\ref{eq:K_bounds_1}) on top of (\ref{eq:P_update_reformulated_long}), we obtain
\begin{align}
\begin{split} \label{eq:P_update_reformulated}
\hat{\mathrm{P}}^{xx}_{t+1} 
\geq \lambda_{t} \mathrm{L}_{t+1} \mathrm{J}_{t}
\hat{\mathrm{P}}^{xx}_{t}
\mathrm{J}_{t}^T \mathrm{L}_{t+1}^T ,
\end{split}
\end{align}
where
\begin{align} \label{eq:lambda_sigma}
\lambda_{t} = 
1 \!+\! \frac{1}{\overline{p}_{t} ( \overline{f}_{t} + \overline{g}_{t}\overline{m}_{t} )^2} 
\Big( 
\underline{q}_{t} 
\!+ \underline{g}_{t}^2 \underline{e}_{t} 
\!+\! \frac{\underline{p}_{t+1}^2 \underline{h}_{t+1}^2} {\overline{r}_{t+1}
( 1 + \frac{\overline{p}_{t+1}\overline{h}_{t+1}^2}{\underline{r}_{t+1}} )^2} 
\Big).
\end{align}
Taking the matrix inverse on both sides of (\ref{eq:P_update_reformulated}) and then let $\Pi_{t} = {\hat{\mathrm{P}}_{t}}^{{xx}^{-1}}$, we have
\begin{align} \label{eq:P_update_reformulated_inverse}
\begin{split}
\Pi_{t+1} \leq \frac{1}{\lambda_{t}} \mathrm{L}_{t+1}^{-T}\mathrm{J}_{t}^{-T} \Pi_{t} \mathrm{J}_{t}^{-1}\mathrm{L}_{t+1}^{-1}.
\end{split}
\end{align}
By pre-multiplying and post-multiplying both sides of the inequality (\ref{eq:P_update_reformulated_inverse}) with $\mathrm{J}_{t}^T\mathrm{L}_{t+1}^T$ and $\mathrm{L}_{t+1}\mathrm{J}_{t}$, respectively, we obtain the result (\ref{eq:prop1_result}) with $\sigma_t = 1 - \frac{1}{\lambda_t}$.
\end{proof}

\begin{proposition} \label{proposition2}
Under the assumptions in Theorem \ref{theorem1}, there exist a positive polynomial functions $\varphi_{t}$ with strictly positive coefficients, for which $\Pi_{t} = {\hat{\mathrm{P}}_{t}}^{{xx}^{-1}}$ satisfies
\begin{align} 
\begin{split} \label{eq:prop2_result}
& \mean \left[ r_{t}^T ( 2 \Pi_{t+1} \mathrm{L}_{t+1} \mathrm{J}_{t} \tilde{x}_{t}
+  r_{t}) \,|\, \tilde{x}_{t} \right]
\leq
\varphi_{t} ( \|\tilde{x}_{t}\| , \delta_{w}, \delta_{\varepsilon}) 
\end{split}
\end{align}
where $r_{t}$ is defined in (\ref{eq:r_t}).
\end{proposition}

\begin{proof}
By substituting (\ref{eq:post_u_error}) into (\ref{eq:assump_phi}) and applying (\ref{eq:assump_FGMH}), (\ref{eq:assump_tau}), we have
\begin{align}
\begin{split} \label{eq:phi_expand}
&\| {r_f}(\tilde{x}_{t},\tilde{u}_{t}) \| 
\leq \kappa^{f}_{t} \big( \| \tilde{x}_{t} \|^2 + \| \mathrm{M}_{t} \tilde{x}_{t} + {r_\phi}(\tilde{x}_{t}) + \varepsilon_{t} \|^2 \big) \\
&\leq \kappa^{f}_{t} \big( \| \tilde{x}_{t} \|^2 + (\| \mathrm{M}_{t} \| \| \tilde{x}_{t} \|)^2 + \| {r_\phi}(\tilde{x}_{t}) \|^2 + \| \varepsilon_{t} \|^2 \big) \\
&\leq \kappa^{f}_{t} \big( ( 1 + \overline{m}_{t}^2 ) \| \tilde{x}_{t} \|^2 + \kappa^{\phi2}_{t} \| \tilde{x}_{t} \|^4 + \| \varepsilon_{t} \|^2 \big).
\end{split}
\end{align}
where we have used the triangle inequality $\| x + y \| \leq \| x \| + \| y \|$ and the induced matrix norm property $\| Ax \| \leq \| A \| + \| x \|$ to obtain the second inequality.
Similarly, substituting (\ref{eq:prior_x_error}) into (\ref{eq:assump_varphi}) and applying (\ref{eq:assump_FGMH}) and (\ref{eq:phi_expand}), we obtain
\begin{align}
\begin{split} \label{eq:varphi_expand}
&\| {r_h}(\tilde{x}_{t+1}^-) \| \\
&\leq \kappa^{h}_{t} \, \| \mathrm{J}_{t} \tilde{x}_{t} + {r_f}(\tilde{x}_{t},\tilde{u}_{t})
+ \mathrm{G}_{t} ({r_\phi}(\tilde{x}_{t}) + \varepsilon_{t}) + w_{t} \|^2 \\
&\leq \kappa^{h}_{t} \Big( (\overline{f}_{t} + \overline{g}_{t}\overline{m}_{t})^2 \| \tilde{x}_{t} \|^2 +\! \big( \kappa^{f2}_{t} (1 + \overline{m}_{t}^2)^2 \!+ \kappa^{\phi2}_{t} \overline{g}_{t}^2 \big) \| \tilde{x}_{t} \|^4 \\
&\qquad\;\;\,\,+ \kappa^{f2}_{t} \kappa^{\phi4}_{t} \| \tilde{x}_{t} \|^8 
+ \overline{g}_{t}^2 \| \varepsilon_{t} \|^2 + \kappa^{f2}_{t} \| \varepsilon_{t} \| ^4
+ \| w_{t} \|^2 \Big).
\end{split}
\end{align}

In addition, we can apply (\ref{eq:K_bounds_1}) on (\ref{eq:r_t}) to obtain the following upper bound of the nonlinear terms:
\begin{align}
\begin{split} \label{eq:r_t_bound}
\| r_{t} \| 
\leq&\; \big( 1 + \frac{\overline{p}_{t+1}\overline{h}_{t+1}^2}{\underline{r}_{t+1}} \big) \| {r_f}(\tilde{x}_{t},\tilde{u}_{t}) \|  
+ \overline{g}_{t} \|{r_\phi}(\tilde{x}_{t})\| \\
&\;+ \frac{\overline{p}_{t+1}\overline{h}_{t+1}}{\underline{r}_{t+1}}
\| {r_h}(\tilde{x}_{t+1}^-) \|,
\end{split}
\end{align}
Besides that, we also have
\begin{align}
\begin{split} \label{eq:r_t_multiply}
&r_{t}^T \left( 2\, \Pi_{t+1} \mathrm{L}_{t+1} \mathrm{J}_{t} \tilde{x}_{t}
+  r_{t} \right) \\
&\leq \|r_{t}\| \Big( \frac{2 (\overline{f}_{t} + \overline{g}_{t}\overline{\phi}_{t})}{\underline{p}_{t+1}} 
\big( 1 + \frac{\overline{p}_{t+1}\overline{h}_{t+1}^2}{\underline{r}_{t+1}} \big) \| \tilde{x}_{t} \|
+ \|r_{t}\| \Big).
\end{split}
\end{align}
Applying (\ref{eq:assump_tau}), (\ref{eq:phi_expand})-(\ref{eq:varphi_expand}) on (\ref{eq:r_t_bound}), 
and then substitute the result into (\ref{eq:r_t_multiply}), we obtain
\begin{align} \label{eq:r_t_polynomial}
r_{t}^T ( 2 \Pi_{t+1} \mathrm{L}_{t+1} \mathrm{J}_{t} \tilde{x}_{t}
+  r_{t} ) 
\leq 
\varphi_{t} (\|\tilde{x}_{t}\|, \! \|w_{t}\|^2, \! \|\varepsilon_{t}\|^2 ),
\end{align}
where $\varphi_{t}$ is a positive polynomial function with indeterminates $(\|\tilde{x}_{t}\|^2, \|w_{t}\|^2, \|\varepsilon_{t}\|^2 )$.
Finally, from the assumptions in (\ref{eq:noise_cov}), we have the following:
\begin{align}
\begin{split} \label{eq:noise_delta}
&\mean[\|w_{t}\|^2] = \tr\left(\mean[w_{t}w_{t}^T]\right) \leq n \delta_w, \\
&\mean[\|\varepsilon_{t}\|^2] = \tr\left( \mean[\varepsilon_{t}\varepsilon_{t}^T] \right) \leq d \delta_\varepsilon, \\
&\mean[\|v_{t}\|^2] = \tr\left(\mean[v_{t}v_{t}^T]\right) \leq m \delta_v.
\end{split}
\end{align}
Considering the fact that even moments of Gaussian noise are multiples of the variance; take the conditional expectation of (\ref{eq:r_t_polynomial}) w.r.t. $\tilde{x}_{t}$ and apply (\ref{eq:noise_delta}), we obtain the result (\ref{eq:prop2_result}).
\end{proof}

\begin{proposition} \label{proposition3}
Under assumptions in Theorem \ref{theorem1}, there exist positive real numbers $c_{w_{t}}, c_{\varepsilon_{t}}, c_{v_{t}}$, where $\Pi_{t} = {\hat{\mathrm{P}}_{t}}^{{xx}^{-1}}$ satisfies
\begin{align}
\begin{split} \label{eq:prop3_result}
\mean\left[ s_{t}^T \Pi_{t+1} s_{t} \,|\, \tilde{x}_{t}\right] \leq c_{w_{t}}\delta_w + c_{\varepsilon_{t}}\delta_\varepsilon + c_{v_{t}}\delta_v,
\end{split}
\end{align}
where $s_{t}$ is defined in (\ref{eq:s_t}).
\end{proposition}

\begin{proof}
Expand $s_{t}^T \Pi_{t+1} s_{t}$ using (\ref{eq:s_t}) to get
\begin{align} 
\begin{split} \label{eq:prop3_expand}
s_{t}^T \Pi_{t+1} s_{t}
=\;& (w_{t} + \mathrm{G}_{t} \varepsilon_{t})^T 
\mathrm{L}_{t+1}^T
\Pi_{t+1} 
\mathrm{L}_{t+1}
(w_{t} + \mathrm{G}_{t} \varepsilon_{t}) \\
&+ v_{t+1}^T\mathrm{K}_{t+1}^T\Pi_{t+1}\mathrm{K}_{t+1}v_{t+1}.
\end{split}
\end{align}
Applying (\ref{eq:assump_sys}) on (\ref{eq:Kalman_linear}) yields the following upper bounds:
\begin{align} \label{eq:K_bounds_2}
\begin{split}  
&\mathrm{K}_{t}^T\Pi_{t}\mathrm{K}_{t}
= \mathrm{R}_{t}^{-1} \mathrm{H}_{t} {\hat{\mathrm{P}}_{t}}^{xx} \mathrm{H}_{t}^T \mathrm{R}_{t}^{-1}
\leq \frac{\overline{p}_{t}\overline{h}_{t}^2}{\underline{r}_{t}^2} \mathrm{I}_m,
\\
&\mathrm{L}_{t}^T\Pi_{t}\mathrm{L}_{t}
= \Pi_{t} - 2 \mathrm{H}_{t}^T \mathrm{R}_{t}^{-1} \mathrm{H}_{t} 
+ \mathrm{H}_{t}^T \mathrm{K}_{t}^T\Pi_{t}\mathrm{K}_{t} \mathrm{H}_{t}
\leq \gamma_{t} \mathrm{I}_n, \\
\end{split}
\end{align}
where
$\gamma_{t} = \frac{1}{\underline{p}_{t}} + \frac{2 \overline{h}_{t}^2}{\underline{r}_{t}} + \frac{\overline{p}_{t}\overline{h}_{t}^4}{\underline{r}_{t}^2}$.
Taking the conditional expectation of (\ref{eq:prop3_expand}) w.r.t. $\tilde{x}_{t}$ and applying (\ref{eq:K_bounds_2}), we have
\begin{align}
\begin{split} \label{eq:noise_expand}
\mean\!\left[ s_{t}^T \Pi_{t+1} s_{t} \,|\, \tilde{x}_{t}\right]
\leq\, & 
\gamma_{t}
( \|w_{t}\|^2 + \overline{g}_{t}^2 \|\varepsilon_{t}\|^2 )
+ \frac{\overline{p}_{t+1}\overline{h}_{t+1}^2}{\underline{r}_{t+1}^2} \|v_{t}\|^2,
\end{split}
\end{align}
where the correlations between the mutually uncorrelated noises vanish.
Substituting (\ref{eq:noise_delta}) into (\ref{eq:noise_expand}), we obtain the result (\ref{eq:prop3_result}) with 
$c_{w_{t}} = 
n \gamma_{t}$, 
$c_{\varepsilon_{t}} = d \overline{g}_{t}^2 \gamma_{t}$, 
$c_{v_{t}} = m \frac{\overline{p}_{t+1}\overline{h}_{t+1}^2}{\underline{r}_{t+1}^2}$.
\end{proof}

We now proceed to the second stage for the proof of Theorem \ref{theorem1}.
Choose the Lyapunov function $V_{t}\left(\tilde{x}_{t}\right) = \tilde{x}_{t}^T \Pi_{t} \tilde{x}_{t}$ with $\Pi_{t} = {\hat{\mathrm{P}}_{t}}^{-1}$, so that from (\ref{eq:assump_P}) we have
\begin{align} \label{eq:lyapunov_bound}
\frac{1}{\overline{p}_{t}}\| \tilde{x}_{t} \|^2 \leq V_{t}\left(\tilde{x}_{t}\right) \leq \frac{1}{\underline{p}_{t}}\| \tilde{x}_{t} \|^2.
\end{align}
Expanding $V_{t+1}\left(\tilde{x}_{t+1}\right)$ using (\ref{eq:x_posterior_error}) gives
\begin{align*} 
\begin{split}
& V_{t+1}(\tilde{x}_{t+1}) = \tilde{x}_{t+1}^T \Pi_{t+1} \tilde{x}_{t+1} \\
&= \begin{pmatrix} \mathrm{L}_{t+1} \mathrm{J}_{t} \tilde{x}_{t} + r_{t} + s_{t} \end{pmatrix}^T 
\Pi_{t+1} 
\begin{pmatrix} \mathrm{L}_{t+1} \mathrm{J}_{t} \tilde{x}_{t} + r_{t} + s_{t} \end{pmatrix} \\ 
&= \tilde{x}_{t}^T \mathrm{J}_{t}^T \mathrm{L}_{t+1}^T 
\Pi_{t+1} \mathrm{L}_{t+1} \mathrm{J}_{t} \tilde{x}_{t}
+ r_{t}^T \Pi_{t+1} \begin{pmatrix} 2\, \mathrm{L}_{t+1} \mathrm{J}_{t} \tilde{x}_{t}
+  r_{t} \end{pmatrix} \\
&\quad+ 2\, s_{t}^T \Pi_{t+1} \mathrm{L}_{t+1} \mathrm{J}_{t} \tilde{x}_{t} + 2\, s_{t}^T \Pi_{t+1} r_{t} 
+ s_{t}^T \Pi_{t+1} s_{t}.
\end{split}
\end{align*}
The conditional expectation of $V_{t+1}(\tilde{x}_{t+1})$ w.r.t. $\tilde{x}_{t}$ yields
\begin{align} \label{eq:lyapunov2}
\begin{split}
& \mean\begin{bmatrix} V_{t+1}(\tilde{x}_{t+1}) \,|\, \tilde{x}_{t} \end{bmatrix} \,=\, \mean\begin{bmatrix} \tilde{x}_{t+1}^T \Pi_{t+1} \tilde{x}_{t+1} \,|\, \tilde{x}_{t}\end{bmatrix} \\
&= \tilde{x}_{t}^T \mathrm{J}_{t}^T \mathrm{L}_{t+1}^T
\Pi_{t+1} \mathrm{L}_{t+1} \mathrm{J}_{t} \tilde{x}_{t} 
+ \mean\begin{bmatrix} s_{t}^T \Pi_{t+1} s_{t} \end{bmatrix} \\
&\quad+ \mean\begin{bmatrix} r_{t}^T \Pi_{t+1} \begin{pmatrix} 2 \mathrm{L}_{t+1} \mathrm{J}_{t} \tilde{x}_{t}
+  r_{t} \end{pmatrix} |\, \tilde{x}_{t} \end{bmatrix},
\end{split}
\end{align}
since
$\mean[ s_{t}^T \Pi_{t+1} \mathrm{L}_{t+1} \mathrm{J}_{t} \tilde{x}_{t} \,|\, \tilde{x}_{t} ] = \mean[ s_{t}^T \,|\, \tilde{x}_{t} ] \mean[ \mathrm{L}_{t+1} \mathrm{J}_{t} \tilde{x}_{t} \,|\, \tilde{x}_{t} ] = 0$
of which both the terms $\Pi_{t+1} \mathrm{L}_{t+1}$ and $\tilde{x}_{t}$ are uncorrelated with $s_{t}$ from (\ref{eq:s_t}).
It also follows that
$\mean[ s_{t}^T \Pi_{t+1} r_{t} \,|\, \tilde{x}_{t} ] = 0$
since the odd moments of Gaussian noise are zero.
Substituting the results of Proposition \ref{proposition1}, \ref{proposition2} and \ref{proposition3} into (\ref{eq:lyapunov2}) yields
\begin{align}
\begin{split} \label{eq:lyapunov3}
&\mean\big[ V_{t+1}\left(\tilde{x}_{t+1}\right) \,|\, \tilde{x}_{t} \big] - V_{t}\left(\tilde{x}_{t}\right) \\
&\leq - \sigma_{t} V_{t}\left(\tilde{x}_{t}\right) + 
\varphi_{t} \left( \|\tilde{x}_{t}\|, \delta_{w}, \delta_{\varepsilon} \right) 
+ \mu_{t} \left( \delta_{w}, \delta_{\varepsilon}, \delta_v \right) ,
\end{split}
\end{align}
where $\mu_{t} = c_{w_{t}}\delta_w + c_{\varepsilon_{t}}\delta_\varepsilon + c_{v_{t}}\delta_v$ gathers the constant terms.

Subsequently, consider
$\underline{\sigma} = {\inf_{t}}\; {\sigma}_{t}$, 
$\overline{p} = {\sup_{t}}\; \overline{p}_{t}$, 
$\overline{\varphi} (\|\tilde{x}_{t}\|,\delta_{\varepsilon}, \delta_v) = {\sup_{t}}\; \varphi_{t} (\|\tilde{x}_{t}\|,\delta_{\varepsilon}, \delta_v)$.
Let $\epsilon$ be the positive root of $\overline{\varphi}(z,\delta_{\varepsilon}, \delta_v) - \frac{\eta \underline{\sigma}}{\overline{p}} z^2$ with $0 < \eta < 1$, and we have
\begin{align}
\begin{split} \label{eq:eta_epsilon}
&\varphi_{t} (\|\tilde{x}_{t}\|, \delta_{\varepsilon}, \delta_v) \\
&\leq \overline{\varphi} (\|\tilde{x}_{t}\|, \delta_{\varepsilon}, \delta_v)
\leq \frac{\eta \underline{\sigma}}{\overline{p}} \|\tilde{x}_{t}\|^2 
\leq \frac{\eta \sigma_{t}}{\overline{p}_{t}} \|\tilde{x}_{t}\|^2 
\leq \eta\sigma_{t} V_{t}(\tilde{x}_{t})
\end{split}
\end{align}
for $\|\tilde{x}_{t}\| \leq \epsilon$, where $\epsilon$ depends on the choices of $(\eta, \delta_{\varepsilon}, \delta_v)$, and the last inequality of (\ref{eq:eta_epsilon}) is due to (\ref{eq:lyapunov_bound}).
Finally, substituting (\ref{eq:eta_epsilon}) into (\ref{eq:lyapunov3}) yields
\begin{align}
\begin{split} \label{eq:final_lyapunov}
& \mean\big[ V_{t+1}\left(\tilde{x}_{t+1}\right) \,|\, \tilde{x}_{t} \big] - V_{t}\left(\tilde{x}_{t}\right)
\leq - \left( 1 - \eta \right) \sigma_{t} V_{t}\left(\tilde{x}_{t}\right)
+ \mu_{t}.
\end{split}
\end{align}
Hence, the inequalities (\ref{eq:final_lyapunov}) and (\ref{eq:lyapunov_bound}) satisfy Lemma \ref{lemma3} with 
$\upsilon_0 = \frac{1}{\underline{p}_{0}}$ and $\upsilon_{t} = \frac{1}{\overline{p}_{t}}$, which proves the exponential boundedness of the posterior state error $\tilde{x}_{t}$ as stated in Theorem \ref{theorem1}. 
Furthermore, the exponential boundedness of the UI error $\tilde{u}_{t}$ follows from (\ref{eq:post_u_error}) and the fact that the covariance of $\varepsilon_{t}$ is bounded via (\ref{eq:noise_cov}).
To prevent the noise term $\mu_{t}$ from driving $\|\tilde{x}_{t}\| \geq \epsilon$, we choose $(\delta_w, \delta_\varepsilon, \delta_v)$ such that
\begin{align} 
\begin{split} \label{eq:exponential_convergence}
\mu_{t} (\delta_w, \delta_\varepsilon, \delta_v)
\leq \frac{\left( 1 - \eta \right) \underline{\sigma}} {\overline{p}} \, \tilde{\epsilon}^2
\leq \left( 1 - \eta \right) \sigma_{t} V_{t}(\tilde{x}_{t})
\end{split}
\end{align}
for some $\tilde{\epsilon} < \epsilon$. 
Substituting (\ref{eq:exponential_convergence}) into (\ref{eq:final_lyapunov}), we have
$
\mean\big[ V_{t+1}\left(\tilde{x}_{t+1}\right) \,|\, \tilde{x}_{t} \big] - V_{t}\left(\tilde{x}_{t}\right) \leq 0
$ which drives $\|\tilde{x}_{t}\|$ towards 0 whenever $\|\tilde{x}_{t}\| \geq \tilde{\epsilon}$.
\end{proof}

\newtheorem*{remark}{Remarks}
\begin{remark} \leavevmode \label{remark:theorem1}
\begin{itemize}
\item 
The assumptions in (\ref{eq:assump_sys}) are standard in nonlinear filter analysis \cite{reif2,xiong,xu}.
The existence of (\ref{eq:assump_P}) depends on the observability of system (\ref{eq:ss_disc}); related discussions can be found in \cite{reif2} therein.
An implication on the remainder errors (\ref{eq:assump_nonl}) is discussed in Section \ref{ssect:RemainderResidual}.
A time-varying state error bound (\ref{eq:lemma3_result}) can be obtained in an online fashion, as demonstrated in Section \ref{ssect:OERB}.
\item
The proposed joint sigma-point transformation scheme (\ref{eq:sgms_posterior})-(\ref{eq:Pcovariance_predict}) incorporates joint state-UI uncertainties in the form of joint covariance $\mathrm{\hat{P}}^{xxuu}_{t}$. This gives rise to the additional terms $\mathrm{G}_{t} \mathrm{M}_{t} \mathrm{\hat{P}}^{xx}_{t} \mathrm{M}_{t}^T \mathrm{G}_{t}^T$ and $\mathrm{G}_{t} \mathrm{E}_{t} \mathrm{G}_{t}^T$ in (\ref{Pxx-_prior_estimate}), which is the linear counterpart of the proposed sigma-point scheme after neglecting remainder errors.
\item
The assumption of $\varepsilon_{t}$ being Gaussian in the UI optimization (\ref{eq:ui_opt}) or model (\ref{eq:ui_model}) can be restrictive when there is deterministic error $\rho_{t}$ arising from modelling assumption or non-convex optimization. 
These deterministic errors can be incorporated into the stability analysis by adding $\rho_{t}$ to $\varepsilon_t$ and determining the upper bound $\|\rho_{t}\| \leq \overline{\rho}_{t}$.
\item
In the presence of large UI errors $\varepsilon_t$ and $\rho_{t}$, the $\mathrm{E}_{t}$ (equivalently $\underline{e}_{t}$) can be set large, which results in a large $\sigma_{t}$ that will improve the error convergence rate, in respect of Lemma 2.
Nevertheless, a large $\mathrm{E}_{t}$ results in large Kalman gain $\mathrm{K}_{t}$, which prompts
the SPKF-nUI to rely more on ${y}_{t}$ and thus amplifies the measurement noise ${v}_{t}$.
\end{itemize}
\end{remark}

\subsection{Nonlinear Remainder Errors} \label{ssect:RemainderResidual}
This subsection discusses the implication on the nonlinear remainder errors in (\ref{eq:assump_nonl}). In particular, the upper bounds of the remainder errors can be approximated as follows.

If $\tilde{X}_{t}$ is small, the remainder in (\ref{eq:xt_Taylor}) can be approximated as
$
R_{f,\hat{X}_{t}}^2(\tilde{X}_{t}) 
\approx
\sum_{|\alpha| = 2} \frac{\tilde{X}_{t}^{\alpha}}{\alpha!} \partial^\alpha f(\hat{X}_{t})
$, 
which by the Multinomial Theorem
$
\sum_{|\alpha| = k} \frac{x^{\alpha}}{\alpha!} 
= \frac{1}{k!} (\sum_{i} x_i)^k
$ and 
$(\sum_{i} x_i)^2 = \sum_{i,j} (xx^T)_{ij}$, 
can also be written as
\begin{align} \label{eq:xt_remainder}
R_{f,\hat{X}_{t}}^2(\tilde{X}_{t}) 
\approx
\frac{1}{2} \sum_{i,j} ( \tilde{X}_{t}\tilde{X}_{t}^T )_{ij} \, \partial^\alpha f(\hat{X}_{t}),
\end{align}
where $\mathrm{A}_{ij}$ denotes the element in row $j$ and column $k$ of matrix $\mathrm{A}$.
Similarly when $\tilde{{X}}_{i,t}$ is small, the remainder in (\ref{eq:xt-_Taylor}) can be approximated as
$
R_{f,\hat{X}_{t}}^2(\tilde{{X}}_{i,t}) 
\approx
\frac{1}{2} \sum_{i,j} ( \tilde{{X}}_{i,t}\tilde{{X}}_{i,t}^T )_{ij} \, \partial^\alpha f(\hat{X}_{t})
$.
Furthermore, from the definition of sigma-points in (\ref{eq:sigma_points}), we have
\begin{align}
\begin{split} \label{eq:xt-_remainder_summed}
\frac{1}{2(n + a)} \sum_{i=1}^{2n} R_{f,\hat{X}_{t}}^2(\tilde{{X}}_{i,t}) 
&\approx 
\frac{1}{2} \sum_{i,j} \mathrm{\hat{P}}^{xxuu}_{t_{ij}} \, \partial^\alpha f(\hat{X}_{t}).
\end{split}
\end{align}
Substitute (\ref{eq:xt_remainder}) and (\ref{eq:xt-_remainder_summed}) into in (\ref{eq:residual_phi}) yields
\begin{align} \label{eq:residual_phi_expand}
{r_f}(\tilde{x}_{t},\tilde{u}_{t}) &\approx 
\frac{1}{2} \sum_{i,j} ( \tilde{X}_{t}\tilde{X}_{t}^T - \mathrm{\hat{P}}^{xxuu}_{t} )_{ij} \, \partial^\alpha f(\hat{X}_{t}).
\end{align}
Given that $\|{\partial}^{\alpha}f(\hat{X}_{t})\| \leq \beta_{f}$ at each $\alpha$ of $\left|\alpha\right| = 2$, we obtain the upper bound of $r_f$ as follows:
\begin{align} \label{eq:phi_bound_norm}
\begin{split}
&\| {r_f}(\tilde{x}_{t},\tilde{u}_{t}) \| 
\;\leq\; \frac{1}{2} \beta_{f} \sum_{i,j} \big| \tilde{X}_{t}\tilde{X}_{t}^T - \mathrm{\hat{P}}^{xxuu}_{t} \big|_{ij} \\
&\leq \frac{1}{2} \beta_{f} \sum_{i,j} \big| \mathrm{I}_{n+m} - \mathrm{\hat{P}}^{xxuu}_{t} (\tilde{X}_{t}\tilde{X}_{t}^T)^{-1} \big|_{ij} \big| \tilde{X}_{t}\tilde{X}_{t}^T \big|_{ij} \\
&\leq \frac{1}{2} \beta_{f} c^{f}_{t} \sum_{i,j} \big| \tilde{X}_{t}\tilde{X}_{t}^T \big|_{ij}
\;\leq\; \frac{1}{2} c^{f}_{t} \beta_{f} \|\tilde{X}_{t}\|_1^2  \\
&\leq \frac{1}{2} c^{f}_{t} \beta_{f} (n+l) \|\tilde{X}_{t}\|_2^2 = \frac{1}{2} c^{f}_{t} \beta_{f} (n+l) \big( \|\tilde{x}_{t}\|_2^2 \!+\! \|\tilde{u}_{t}\|_2^2 \big)
\end{split}
\end{align}
with
$
c^{f}_{t} = \| \mathrm{I}_{n+m} - \mathrm{\hat{P}}^{xxuu}_{t} (\tilde{X}_{t}\tilde{X}_{t}^T)^{-1} \|_{\max}
$, where
$
\| A \|_{\max} = \max_{i,j} |A|_{ij}
$.
The fifth inequality of (\ref{eq:phi_bound_norm}) uses the equivalence of norms, $\|x\|_1 \leq \sqrt{n} \|x\|_2$, $x \in \mathbb{R}^n$.
Hence, we obtain $\kappa^{f}_{t} \approx c^{f}_{t} \beta_{f} (n+l)$ in (\ref{eq:assump_phi}).
Similarly, we can apply the same procedures (\ref{eq:xt_remainder})-(\ref{eq:phi_bound_norm}) to ${r_\phi}$ (\ref{eq:residual_tau}) and ${r_h}$ (\ref{eq:residual_varphi}) to approximate $\kappa^{\phi}_{t}$ in (\ref{eq:assump_tau}) and $\kappa^{h}_{t}$ in (\ref{eq:assump_varphi}), respectively.

The term $\beta_{f}$ in (\ref{eq:phi_bound_norm}) can be obtained as spectral norm of the Hessian of state model $f$,
$\beta_{f} = {\max}_{1 \leq i \leq n} \, {\sup}_{X \in S} \; \frac{\partial f_i}{\partial X}$ \cite{reif2}.
When $\tilde{X}_{t}\tilde{X}_{t}^T \approx \mathrm{\hat{P}}^{xxuu}_{t}$, the coefficient $c^{f}_{t}$ in (\ref{eq:phi_bound_norm}) becomes small and regulates the remainder error ${r_f}$. The same implication can be made for ${r_\phi}$ and ${r_h}$.
Consequently, these regularizations exclusive to the SPKF, facilitate small $\overline{\varphi}$ and $\eta$ in (\ref{eq:eta_epsilon}) and results in a smaller error bound and a faster error convergence, when compared to the EKF. These advantages of the SPKF are illustrated in Section \ref{sect:ResultsandDiscussions}.
\section{Illustrative Examples} \label{sect:IllustrativeExample}
In this section, we present two case studies to demonstrate the proposed SPKF-nUI.
The first case study is a simulation-based rigid-link robot that exhibits trigonometric nonlinearity. It is conducted using a systematic square-wave input to verify the convergence of SPKF-nUI guaranteed by Theorem \ref{theorem1}. This case study uses an analytical model with a nonlinear least-squares UI optimization (\ref{eq:ui_opt}).
The second case study is a physical soft robot, i.e., robot made of soft materials which is known to exhibit rich and nonlinear dynamics \cite{junn_soro}. The robot is actuated using both gradual-oscillatory and fast-switching randomized inputs to cover a wide range of complex dynamics. Consider that analytical modelling is challenging for soft robots, we identify the system models and nonlinear UI model (\ref{eq:ui_model}) empirically using deep learning.
Lastly, we detail the process of obtaining the time-varying state error bounds.

\subsection{Case Study 1: Rigid-link robot} \label{ssect:SAM}

\begin{figure}[t]
\centering
\includegraphics[width=0.55\columnwidth]{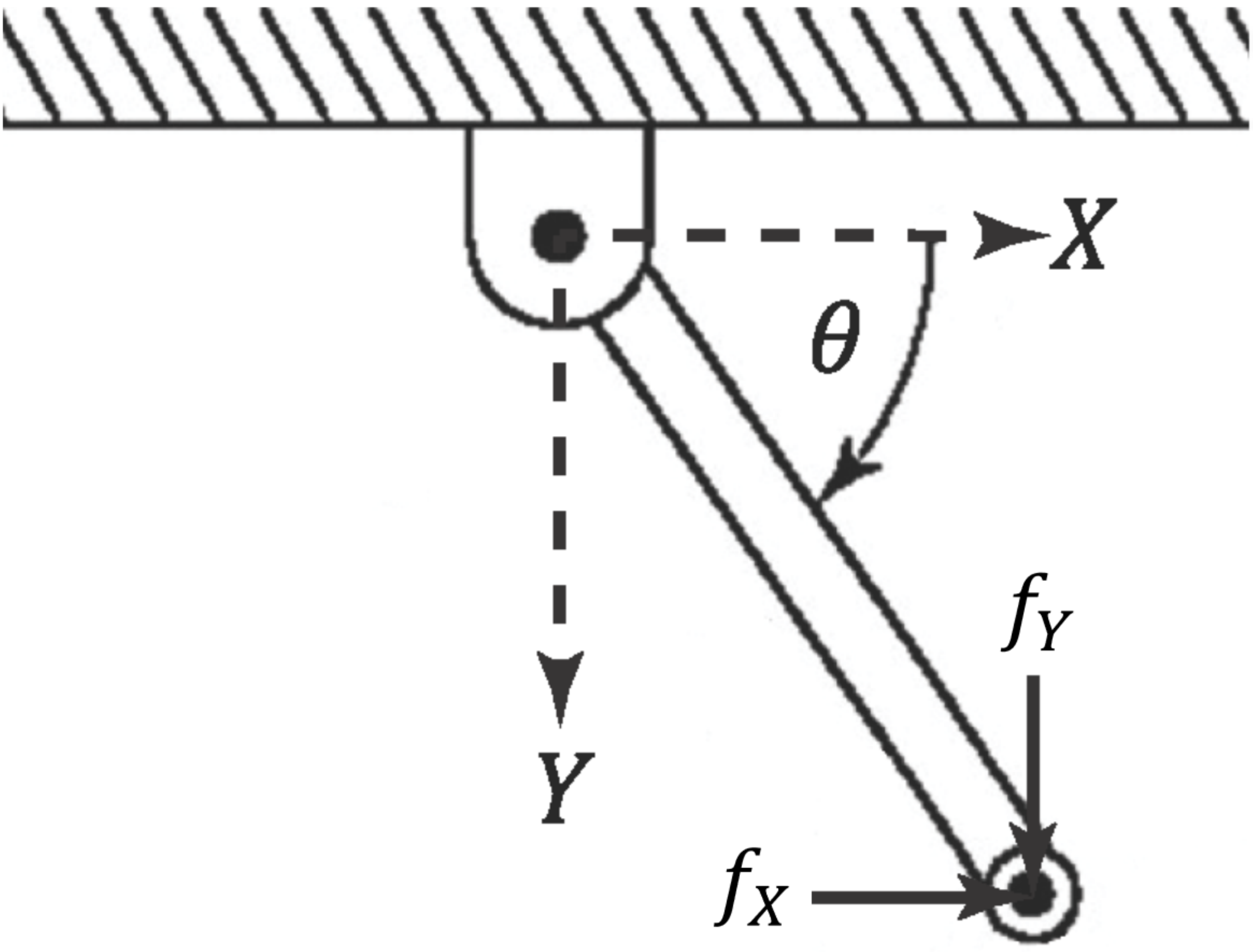}
\caption{\textbf{Case study 1: Rigid-link robot.} ${\theta}$ is the angular displacement, and the two-axis force $f = \begin{bmatrix} f_X & f_Y \end{bmatrix}^T$ is applied at the link tip.}
\label{fig:SLRM}
\end{figure}

The first case study is a rigid-link robot (Fig. \ref{fig:SLRM}) where $\theta$ is the link angular displacement ($\dot{\theta}$ and $\ddot{\theta}$ denote the velocity and acceleration, respectively), and $f_{X}$, $f_{Y}$ are the horizontal and vertical forces acting on the tip. The equation of motion (EoM) of the robot can be derived using iterative Newton-Euler dynamics. Denote the state $x$ as $(x_1, x_2) = (\dot{\theta}, \theta)$, and the UI $u$ as $(u_1, u_2) = (f_{X}, f_{Y})$. Then from \cite[Chapter~6]{Craig2022}, the equation of motion (EoM) of the robot is given by:
\begin{align}
\begin{split} \label{eq:EoM}
ml^2 \ddot{\theta} &= \Phi^{\mathrm{EoM}} (x,u) \\
&= - b \, x_1 + m g \, l \cos x_2  + u_1 \, l \sin x_2 - u_2 \, l \cos x_2
\end{split}
\end{align}
where $m$, $l$, $b$ and $g$ respectively are the link mass, link length, damping coefficient and gravitational acceleration.
A full derivation of (\ref{eq:EoM}) is provided in supplementary materials.

By substituting the angular acceleration $\ddot{\theta}_t$ at time $t$, from (\ref{eq:EoM}) into
the following Euler integration equations, $x_{1_{t+1}} = x_{1_{t}} + \ddot{\theta}_{t} h$ and $x_{2_{t+1}} = x_{2_{t}} + x_{1_{t}} h + \frac{1}{2} \ddot{\theta}_{t} {h}^2$ with step size $h$, a state-space representation as (\ref{eq:ss_disc}) is obtained as follows:
\begin{align}
\label{eq:ss_f}
\begin{bmatrix} x_{1_{t+1}} \\ x_{2_{t+1}} \end{bmatrix} 
&= 
\begin{bmatrix} 1 & 0 \\ h & 1 \end{bmatrix} \begin{bmatrix} x_{1_{t}} \\ x_{2_{t}} \end{bmatrix} 
+ \begin{bmatrix} h \\ \frac{1}{2}{h}^2 \end{bmatrix} 
\Phi^{\mathrm{EoM}}(x_{t},u_{t}), \\
\label{eq:ss_h}
y_{t} &= \begin{bmatrix} x_{1_{t}} & l \cos x_{2_{t}} & l \sin x_{2_{t}} \end{bmatrix}^T,
\end{align}
where we have also modeled the angular velocity and Cartesian coordinates of the link tip position as measurements in (\ref{eq:ss_h}).
Note that the UIs are not linearly separable from the states in the state model (\ref{eq:ss_f}). 

To obtain an analytical expression of the function $\Phi$ for UI optimization (\ref{eq:ui_opt}), in this case study, we leverage the quasi-static approximation \cite{junn_soro}, a widely employed robot modeling technique that assumes the robot is momentarily in equilibrium with zero acceleration, i.e., $\ddot{\theta}_{t} = 0$ at each time $t$. This is equivalent to imposing a zero-order hold $x_{1_{t+1}} = x_{1_{t}}$ on the first system state in (\ref{eq:ss_f}), which is reasonable when rapid dynamics are absent, or when the dynamics change slowly relative to the control input. 
Applying the quasi-static approximation to the EoM in (\ref{eq:EoM}) yields $\Phi^{\mathrm{EoM}} = 0$, which can be solved (and $u_{t}$ can be estimated) by reformulating it as the following nonlinear least-squares problem that minimizes the squared Euclidean distance $\| \Phi^{\mathrm{EoM}}(x_{t},u_{t}) - 0 \|$ w.r.t. $u_{t}$:
\begin{align}
\begin{split} \label{eq:Psi}
u_{t} &= \arg \min_{u_t} \left\| \Phi^{\mathrm{EoM}}(x_{t},u_{t}) \right\|^2.
\end{split}
\end{align}
Therefore, $\Phi^{\mathrm{EoM}}$ corresponds to the nonlinear residual $\Phi$ in (\ref{eq:ui_opt}), and in this context, the UI noise $\varepsilon_{t}$ represents optimization error arising from the quasi-static assumption. 
Here, we solve the optimization (\ref{eq:Psi}) via the conjugate gradient algorithm \cite{chong}.

\begin{figure}[t]
\centering
\includegraphics[width=1\columnwidth]{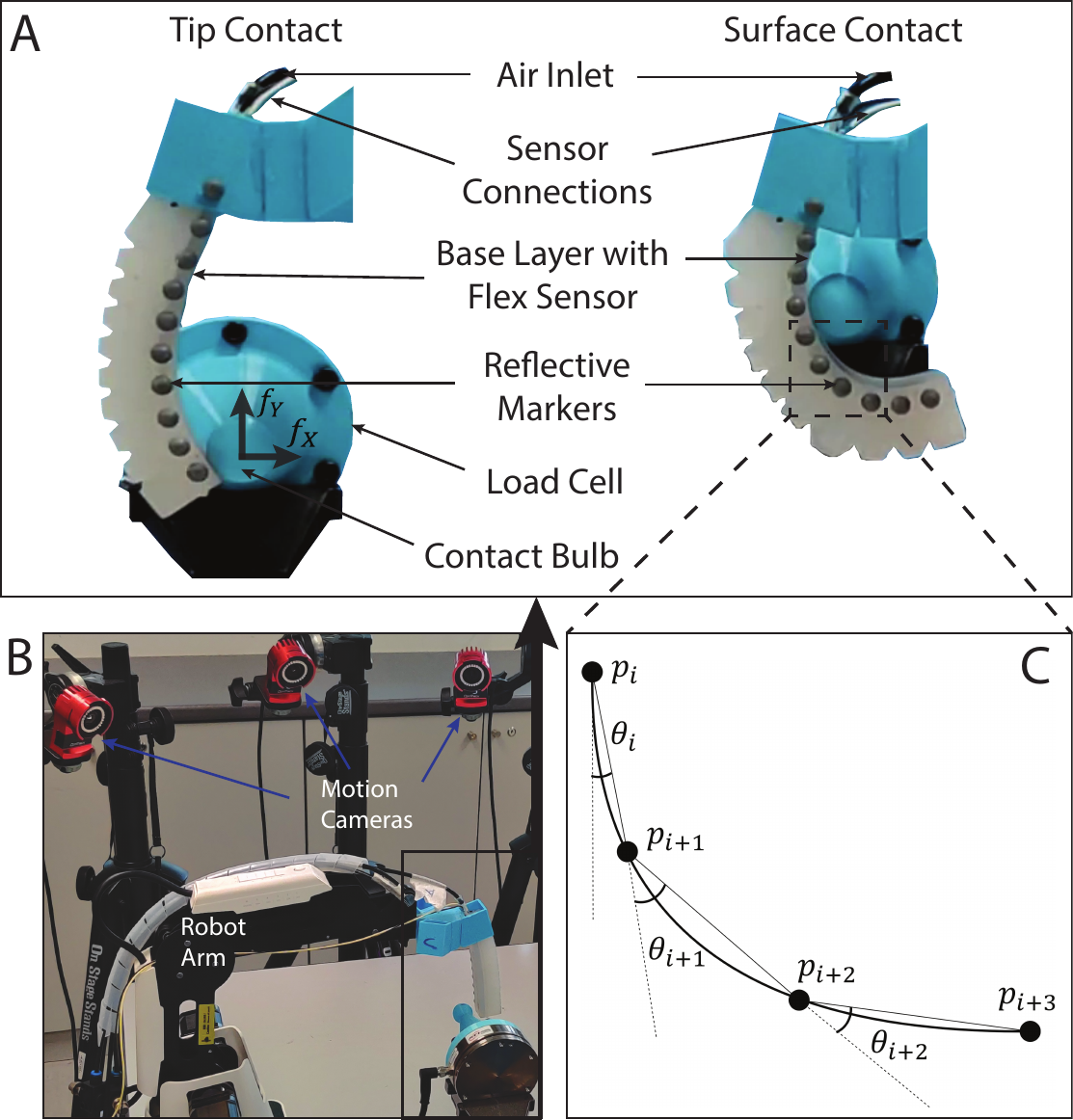}
\caption{\textbf{Case study 2: Pneumatic soft actuator (PSA)}. (A, Left) Using a contact bulb attached on top of a multi axis load cell (Axia80, ATI Industrial Automation Inc.), contact forces are applied on the PSA at its fingertip (i.e., Tip Contact). (A, Right) Contact forces are applied on the PSA along its surface (i.e., Surface Contact). $\begin{bmatrix} f_X & f_Y \end{bmatrix}^T$ is the two-axis reaction force from the PSA (with direction opposite to the contact forces) measured by the load cell attached to the contact bulb. 10 reflective camera markers are placed evenly along the inextensible base layer to capture the PSA motion. 
(B) Three cameras are used to track the marker coordinates. A rigid robot arm is used to maneuver the position of PSA.
(C) Planar line segment model to characterize PSA bending, where $\begin{bmatrix}X_{i} \quad Y_{i}\end{bmatrix}^T$ are the coordinates of the $i^{th}$ camera marker and $\theta_i$ is the $i^{th}$ segmental bending angle.}
\label{fig:scenarios}
\end{figure}

\subsection{Case Study 2: Pneumatic Soft Actuator (PSA)} \label{ssect:PSA}

In fully autonomous systems, perceptive information such as the internal states and the external excitations (UIs) are crucial for informed decision-making in industrial tasks. However, integrating sensors in soft robots to measure these perceptive variables remains an arduous task. 
Soft robots have infinite degrees of freedom which would require a substantial amount of sensors for measuring these perceptive variables. Moreover, integrating many sensors into a soft robot risk altering its mechanical characteristics and functionality \cite{junn_soro}.

In this case study, we consider a physical pneumatic soft actuator (PSA) as shown in Fig. \ref{fig:scenarios}A, and aim to estimate its bending angles (states) and contact forces (UIs).
Four experimental scenarios (titles of Fig. \ref{fig:SR_results}) comprising two robot configurations and two actuation patterns are conducted on the PSA.
In \textit{Tip Contact}, PSA is actuated to perform bending with a contact bulb placed in front of its tip to mimic a surface contact, as shown in Fig. \ref{fig:scenarios}A (Left). The PSA is configured to move in towards and out from the contact bulb along the $X$-axis to covers a range of possible gripper configurations and contact points.
In \textit{Surface Contact}, PSA is configured to randomly move up and down (along $Y$-axis) to allow contact along the whole PSA's front surface, as shown in Fig. \ref{fig:scenarios}A (Right).
For each configuration, we input a gradual oscillatory pressure for \textit{Oscillatory Actuation}, and a faster randomized pressure for \textit{Random Actuation}. 
These experiments simulate grasping motions with complex nonlinear dynamics, critical in validating the efficacy of our proposed filter.
Two separate (training and validation) datasets of the input pressure, flex sensor reading, marker coordinates and contact forces are collected from the experiments at 10 Hz. The marker coordinates are recorded by motion cameras (Fig. \ref{fig:scenarios}B) and converted to segmental bending angles using the line-segment method \cite{junn_soro} (Fig. \ref{fig:scenarios}C). The two-axis contact forces are measured by load cell attached to the contact bulb (Fig. \ref{fig:scenarios}B).

Here, we consider the following probabilistic Gated Recurrent Unit (GRU) \cite{zeyang_tii}, which is a class of stochastic RNNs, for data-driven modelling of the PSA robot system:
\begin{align} \label{eq:GRU}
\begin{split}
&z_{t} = \psi\, \big( W_{z} \!\begin{bmatrix} {x}_{t}^T \!&\! {u}_{t}^T \!&\! {h}_{t}^T \end{bmatrix}^T\! + {b}_{z} \big), \\
&n_{t} = \psi\, \big( W_{n} \!\begin{bmatrix} {x}_{t}^T \!&\! {u}_{t}^T \!&\! {h}_{t}^T \end{bmatrix}^T\! + {b}_{n} \big), \\
&\tilde{h}_{t} = W_{h_1} \!\begin{bmatrix} {x}_{t}^T \!&\! {u}_{t}^T \end{bmatrix}^T\! + {b}_{h_1} + z_t \odot \big( W_{h_2} h_{t} + {b}_{h_2} \big), \\
&(x_{t+1} \,,\, \sigma^{w}_{t}) = \psi\, \big( {W}_{x} h_{t} + {b}_{x} \big), \\
&y_{t} = \psi\, \big( W_{y} \!\begin{bmatrix} {x}_{t}^T \!&\! {h}_{t}^T \end{bmatrix}^T\! + {b}_{y} \big), \\
&(u_{t} \,,\, \sigma^{\varepsilon}_{t}) = \psi\, \big( W_{u} \!\begin{bmatrix} {x}_{t}^T \!&\! {h}_{t}^T \end{bmatrix}^T\! + {b}_{u} \big), \\
&h_{t+1} = n_{t} \odot h_{t} + (1 - {n_t}) \odot \text{tanh}\, \big( \tilde{h}_{t} \big),
\end{split}
\end{align}
where $h_{t}\in \mathbb{R}^{128}$ is the GRU's recurrent hidden states.
Here, $\Theta$ represents the set of NN weight matrices $W$
and bias vectors $b$,
$\psi$ is the sigmoid activation function, and $\odot$ denotes the Hadamard (element-wise) product. Here, the GRU (\ref{eq:GRU}) also predicts the respective standard deviations
$\sigma^{w}_{t}$, $\sigma^{\varepsilon}_{t}$ 
of the isotropic covariance parameters
$\mathrm{Q}_{t}$, $\mathrm{E}_{t}$.
By rendering the hidden states $h_{t}$ implicit, the GRU equations (\ref{eq:GRU}) can also be formulated as follows:
\begin{align}
\begin{split} \label{eq:ss_GRU}
&(\chi_{t+1}, \sigma^{w}_{t}) = f_{{\Theta}_t}(x_{t},u_{t}), \quad
y_{t} = h_{{\Theta}_t}(x_{t}), \\
&(u_{t}, \sigma^{\varepsilon}_{t}) = \phi_{{\Theta}_t}(x_{t}),
\end{split}
\end{align}
where the time-varying models $f_{\Theta_{t}}$, $h_{\Theta_{t}}$, $\phi_{\Theta_{t}}$ thus corresponds to the nonlinear system (\ref{eq:ss_disc}) and UI model (\ref{eq:ui_model}).
Due to the intrinsic nonlinearity of RNNs, existing 
methods \cite{zhengzhao1,zhengzhao2,anagnostou,Zhongjin} that assumed linearly separable UI are thus not applicable to system (\ref{eq:ss_GRU}).
As opposed to the UI optimization in Case Study 1 (Section \ref{ssect:SAM}), we employ a data-driven RNN model $\phi_{\Theta_{t}}$ for the UI prediction in this case study.
The GRU parameters $\Theta$ are trained end-to-end using negative log-likelihoods supervised by the training data.
Finally, using the measurement $y \in \mathbb{R}$ of the single embedded flex sensor reading and the input pressure $d \in \mathbb{R}$, we estimate the states $x = \begin{bmatrix} \theta_1 & \dots & \theta_9 \end{bmatrix}^T \in \mathbb{R}^9$ (i.e., segmental bending angles) and the contact forces $(f_X,f_Y)$ of the UIs $u = \begin{bmatrix} f_X & f_Y & d \end{bmatrix}^T$.
The sampled data of bending angles and the contact forces are used only as ground truths when assessing estimation results on the validation dataset.


\subsection{State Error Bound Computation} \label{ssect:OERB}
To compute the time-varying state error bound in (\ref{eq:lemma3_result}) for the Case Study 1, the system model Jacobians $(\mathrm{F}_{t}, \mathrm{G}_{t}, \mathrm{H}_{t})$ are first obtained via first-order linearization using the filter estimates $(\hat{x}_{t}^-,\hat{x}_{t},\hat{u}_{t})$ at each time-step.
Then, $\mathrm{M}_{t}$ is obtained via the Gauss–Newton method (\ref{eq:jacobian_u}).
Subsequently, we obtain the scalar matrix bounds of the inequalities (\ref{eq:assump_sys}) by computing the largest and smallest singular values, $\sigma^{s}_{\max}(\mathrm{A})$ and $\sigma^{s}_{\min}(\mathrm{A})$, of the system model Jacobians, the covariance parameters $(\mathrm{Q}_{t}, \mathrm{E}_{t}, \mathrm{R}_{t})$, and the posterior state covariance $\mathrm{\hat{P}}^{xx}_{t}$.
Based on these computed scalar matrix bounds, we can then obtain $\lambda_{t}$ from (\ref{eq:lambda_sigma}) and take $1 - \sigma_{t} = \frac{1}{\lambda_{t}}$.

The state error bound analysis in Theorem \ref{theorem1} is particularly useful for detecting $(w,v,\varepsilon)$ and identifying their covariances. 
This can be achieved by computing the state error bound using the right-hand side of (\ref{eq:lemma3_result}) concurrently with the state estimation error $\mean\left[\| \tilde{x}_{t} \|^2\right]$. In the presence of unattributed errors, the state estimation error will exceed the computed error bound.
Given that the exact UI covariance $\mathrm{E}_{t}$ is unknown in the Case Study 1 (Section \ref{ssect:SAM}), we can approximate it with $\mathrm{E}_{t}$ and fine-tune it until the computed error bound constitutes a upper bound of $\mean\left[\| \tilde{x}_{t} \|^2\right]$. 
Fig. \ref{fig:RR_results}D shows the error bounds (ERBs) computed using $\mathrm{E}_{t} = 0 \times \mathrm{I}_{d}$ before tuning and $\mathrm{E}_{t} = 35 \times \mathrm{I}_{d}$ after tuning, respectively.
Error bound analysis is not conducted for Case Study 2 due to the lack of information on the actual values of $\mathrm{Q}_{t}$ and $\mathrm{E}_{t}$.

\section{Results and Discussions} \label{sect:ResultsandDiscussions}
In this section, we present and analyze the state and UI estimation results of the proposed SPKF-nUI against existing baseline filters on the introduced case studies.

\subsection{Baselines}
To evaluate the performance of our proposed SPKF-nUI, we benchmark it against the existing nonlinear state-UI filters, in particular, EKF-UI\cite{Ghahremani}, EKF-MVU\cite{joseph}, SPKF-UI\cite{anagnostou}, SPKF-MVU\cite{zhengzhao1}, and the conventional SPKF\cite{arasaratnam}, EKF without UI estimation. Considering that UIs are not linearly separable from the state model in (\ref{eq:ss_disc}), the (linear) least-squares UI optimization of these state-UI baselines is performed on top of first-order local linearization. Also, following \cite{joseph}, we set ${u}_{t} = 0$ for the first state prediction stage of these baselines. For the more complex Case Study 2, least-squares optimization of the state-UI baselines is replaced by the RNN model $\phi_{\Theta_{t}}$ from (\ref{eq:ss_GRU}) to prevent extreme linearization errors.
To illustrate the importance of the proposed joint sigma-point transformation scheme (\ref{eq:sgms_posterior})-(\ref{eq:Pcovariance_predict}), we also compare against EKF-nUI, an EKF counterpart of our proposed SPKF-nUI. The EKF-nUI updates prior state covariance using (\ref{Pxx-_prior_estimate}), i.e., the linear counterpart of the proposed sigma-point transformation.

In addition, we introduce two filter variants SPKF-nUI-I and SPKF-nUI-II, as well as their EKF counterparts EKF-nUI-I and EKF-nUI-II. In particular, the SPKF-nUI-I and the EKF-nUI-I use prior state sigma-points for UI estimation (\ref{eq:uiopt_posterior}). 
The SPKF-nUI-II and the EKF-nUI-II 
employ the conventional state and covariance updates of the SPKF and the EKF, respectively.
Here, we also include the Crem\'{e}r-Rao Lower Bound (CRLB) for benchmarking the state estimation.

\subsection{Case Study 1: Rigid-link robot}

The simulation parameters are set to 
$m = 1$, $l = 1$, $b = 5$, $g = 9.81$, $h = 0.01$
The initial states are set as
$x_{0} = \begin{bmatrix} 0 & 0 \end{bmatrix}^T$.
The noise signals $w_{t}$ and $v_{t}$ are set to have covariances
$\mean[w_{t}w_{t}^T] = 0.001 \times \mathrm{I}_2$, and an intense 
$\mean[v_{t}v_{t}^T] = 0.5 \times \mathrm{I}_3$.
The UIs are set to be
$u_{1_{t}} = 10 \, \mathrm{sgn} \left( \sin 0.1 \pi t \right)$,  
$u_{2_{t}} = 0$
where $\mathrm{sgn}$ is the signum function.
Using these settings, we conduct 50 Monte Carlo (MC) simulations and obtain 50 $(x_t,u_t,y_t)$ sequences.
All filters are initialized with
$\hat{x}_{0} = \begin{bmatrix} 0 & \frac{\pi}{2} \end{bmatrix}^T$,  
$\mathrm{\hat{P}}_{0} = 0.5 \mathrm{I}_2$.
The filter covariance parameters are set to
$\mathrm{Q}_{t} = \mean[w_{t}w_{t}^T]$,
$\mathrm{R}_{t} = \mean[v_{t}v_{t}^T]$.
The $\mathrm{E}_{t}$ exclusive to SPKF-nUI is set to
$\mathrm{E}_{t} = 35 \times \mathrm{I}_2$, which we obtained via error bound analysis as described in Section (\ref{ssect:OERB}).
Each filter produces 50 estimations of the simulated $(x_t,u_t)$ from the simulated $y_t$.

\begin{figure}[t]
\centering
\includegraphics[width=1\columnwidth]{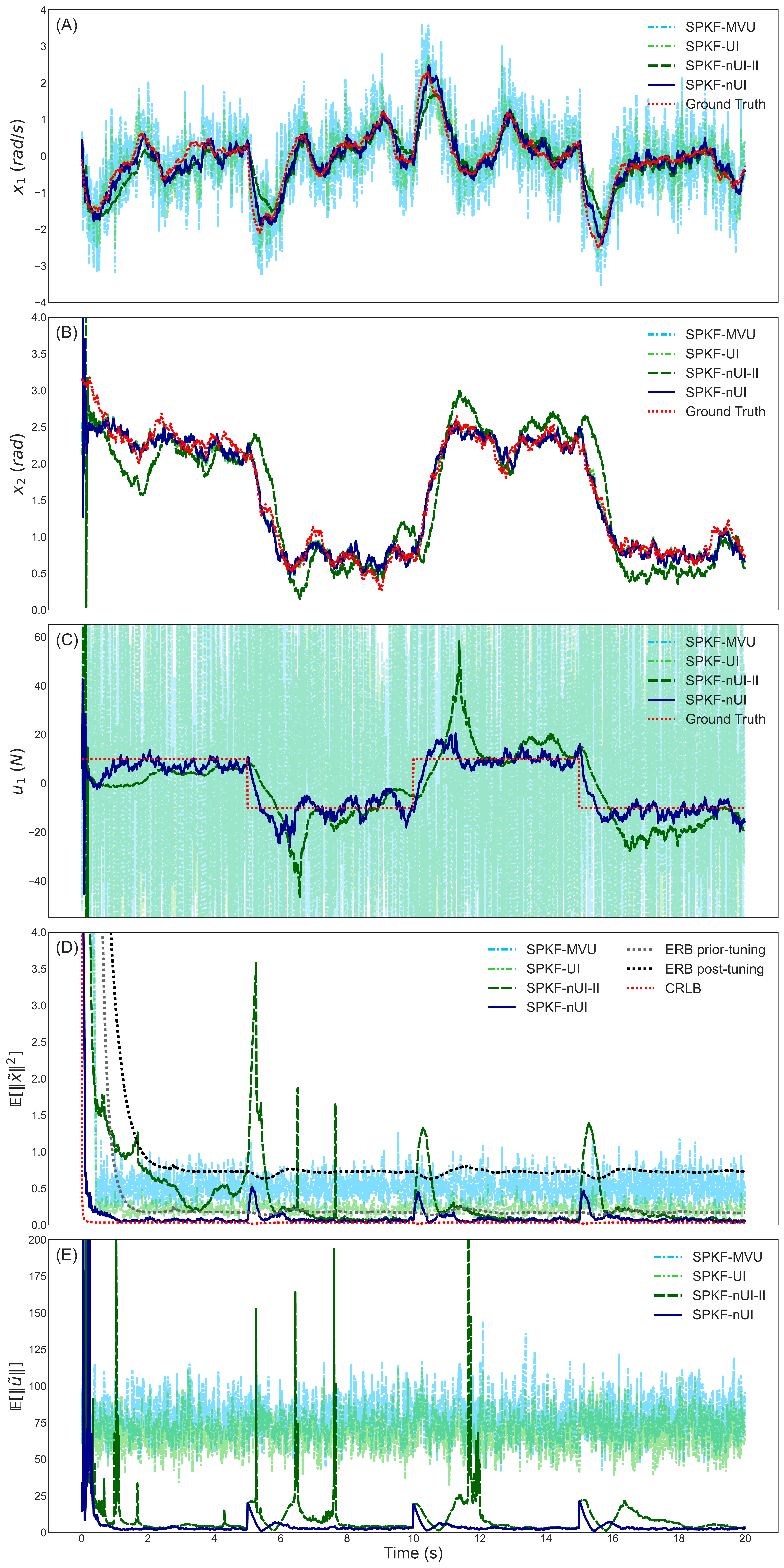}
\caption{\textbf{Estimation results of Case Study 1}.
(A) and (B) show the ground truths $x_{1_t}$, $x_{2_t}$ and estimates $\hat{x}_{1_t}, \hat{x}_{2_t}$ of the states.
(C) shows the $u_{1_t}$ (ground truth) and estimate $\hat{u}_{1_t}$ of the non-zero UI.
These results in (A-C) are obtained from the first MC simulation.
(D) shows the NMSE $\mean[\|\tilde{x}_{t}\|^2]$ of the state estimates, the state error bounds (ERBs) of the SPKF-nUI (Section \ref{ssect:OERB}), and the norm of the theoretical CRLB (benchmark).
(E) shows the NMSE $\mean[\|\tilde{u}_{t}\|]$ of the UI estimates.
These results in (D-E) are obtained via averaging across the 50 MC simulations. 
}
\label{fig:RR_results}
\end{figure}

Table \ref{tab:CaseStudy1} tabulates the normalized mean-square-error (NMSE) and signal-to-noise (SNR) ratio of the estimations, where the proposed SPKF-nUI achieves superior state and UI estimation performances.
Fig. \ref{fig:RR_results} shows the estimations and their NMSEs.
Notice in Fig. \ref{fig:RR_results}D that the state error of EKF-nUI converge slower towards the theoretical CRLB at the beginning when compared to the SPKF-nUI.
This is due to the first-order linearization and a larger nonlinear remainder error in EKF-based estimation as explained in Section \ref{ssect:RemainderResidual}.
Comparing the results of SPKF-nUI and SPKF-nUI-I in Table \ref{tab:CaseStudy1}, we notice that better performance is achieved when UI estimation is performed using posterior state estimates.
Also, notice that all the UI errors in Fig. \ref{fig:RR_results}C and Fig. \ref{fig:RR_results}E are large after the input ramps due to modelling errors arise from the quasi-static assumption.
Compared to SPKF-nUI-II and EKF-nUI-II, our proposed SPKF-nUI and EKF-nUI are able to generate estimates that are more robust to these uncertainties of the UI optimization (\ref{eq:ui_opt}), as shown in Fig. \ref{fig:RR_results}, by virtue of the joint sigma-point transformation scheme (\ref{eq:sgms_posterior})-(\ref{eq:Pcovariance_predict}) and its linear counterpart (\ref{Pxx-_prior_estimate}).
In addition, notice in Fig. \ref{fig:RR_results}C that the UI estimates of SPKF-MVU, EKF-MVU and SPKF-UI, EKF-UI exhibit severe fluctuations, with their NMSEs exceeding the boundary of Fig. \ref{fig:RR_results}E. This is due to large linearization errors introduced via the linear least-squares of these baselines, which have to be performed on top of local linearization given that the UIs are not linearly separable in (\ref{eq:ss_disc}). 
Furthermore, these baseline filters impose large Kalman correction upon their state predictions to compensate for large UI errors. Consequently, this propagates measurement noise and gives rise to noisy state estimates, as shown in Fig. \ref{fig:RR_results}A,D and indicated by the low state SNRs in Table \ref{tab:CaseStudy1}.
The SPKF and EKF perform poorly here due to their inability to estimate UI.

\subsection{Case Study 2: Pneumatic Soft Actuator (PSA)}

We conduct 10 MC simulations of measurement noise $v_t$ sequences with covariance $\mean[v_{t}v_{t}^T] = 1 \times 10^{-3}$ and add them on top of the sampled $y_{t}$ sequences.
All filters are initialized with
$\hat{x}_{0} = 0 \times \mathbf{1}_9$, where $\mathbf{1}_n$ denotes column vector of size $n$ with unit entries, and
$\mathrm{\hat{P}}_{0} = 0.1 \times \mathrm{I}_9$.
The filter covariance parameters are set to
$\mathrm{Q}_{t} = \sigma^{w}_{t} \times \mathrm{I}_9$
and
$\mathrm{R}_{t} = \mean[v_{t}v_{t}^T]$.
The $\mathrm{E}_{t}$ exclusive to the SPKF-nUI is set to
$\mathrm{E}_{t} = \sigma^{\varepsilon}_{t} \times \mathrm{I}_2$.
Since the SPKF and EKF do not estimate UIs, we set
$\hat{u}_{t} = \begin{bmatrix} 0 & 0 \end{bmatrix}^T$
for them.
Each filter produces 10 estimations of the sampled $(x_t,u_t)$ from the $y_t$ with amplified noise.

\begin{table}[t]
\centering
\caption{\textbf{State NMSEs $\mean[ \|\tilde{x}\|^2 ]$ and UI NMSEs $\mean[ \|\tilde{u}\|^2 ]$ of Case Study 1.} NA indicates that result is not available. NC indicates non-converging (very large) errors.}
\label{tab:CaseStudy1}
\resizebox{\columnwidth}{!}{%
\begin{tabular}{@{}lrrcr@{}}
\toprule
\multicolumn{1}{c}{\multirow{2}{*}{Method}} & \multicolumn{2}{c}{System State} & \multicolumn{2}{c}{Unknown Input} \\ \cmidrule(l){2-5} 
\multicolumn{1}{c}{} & \multicolumn{1}{c}{NMSE} & \multicolumn{1}{c}{SNR} & NMSE & \multicolumn{1}{c}{SNR} \\ \midrule
SPKF-nUI & \textbf{0.670 ± 0.116} & 40.5 ± 3.7 & \multicolumn{1}{r}{\textbf{0.598 ± 0.080}} & 33.2 ± 1.5 \\
EKF-nUI & 1.092 ± 1.010 & 41.7 ± 5.4 & \multicolumn{1}{r}{0.667 ± 0.107} & 34.1 ± 1.6 \\ \midrule
SPKF-nUI-I & 0.824 ± 0.174 & 40.1 ± 3.6 & \multicolumn{1}{r}{0.623 ± 0.082} & 33.3 ± 1.5 \\
EKF-nUI-I & 0.998 ± 0.723 & 41.8 ± 5.8 & \multicolumn{1}{r}{0.712 ± 0.114} & 34.8 ± 1.6 \\ \midrule
SPKF-nUI-II & 3.621 ± 4.309 & 35.2 ± 3.2 & \multicolumn{1}{r}{91.8 ± 266.2} & 28.5 ± 16.2 \\
EKF-nUI-II & 3.618 ± 3.640 & 42.4 ± 6.3 & \multicolumn{1}{r}{152.8 ± 511.2} & 31.5 ± 13.7 \\ \midrule
SPKF-MVU & 3.630 ± 1.832 & 26.7 ± 6.0 & \multicolumn{2}{c}{NC} \\
EKF-MVU & 2.650 ± 1.153 & 30.1 ± 5.2 & \multicolumn{2}{c}{NC} \\ \midrule
SPKF-UI & 2.593 ± 1.799 & 29.2 ± 6.1 & \multicolumn{2}{c}{NC} \\
EKF-UI & 2.762 ± 1.412 & 31.1 ± 5.3 & \multicolumn{2}{c}{NC} \\ \midrule
SPKF & 3.280 ± 1.503 & 35.1 ± 5.9 & \multicolumn{2}{c}{NA} \\
EKF & 3.456 ± 0.411 & 45.3 ± 3.9 & \multicolumn{2}{c}{NA} \\ \bottomrule
\end{tabular}%
}
\end{table}

\begin{figure*}[t]
\centering
\includegraphics[width=1\textwidth]{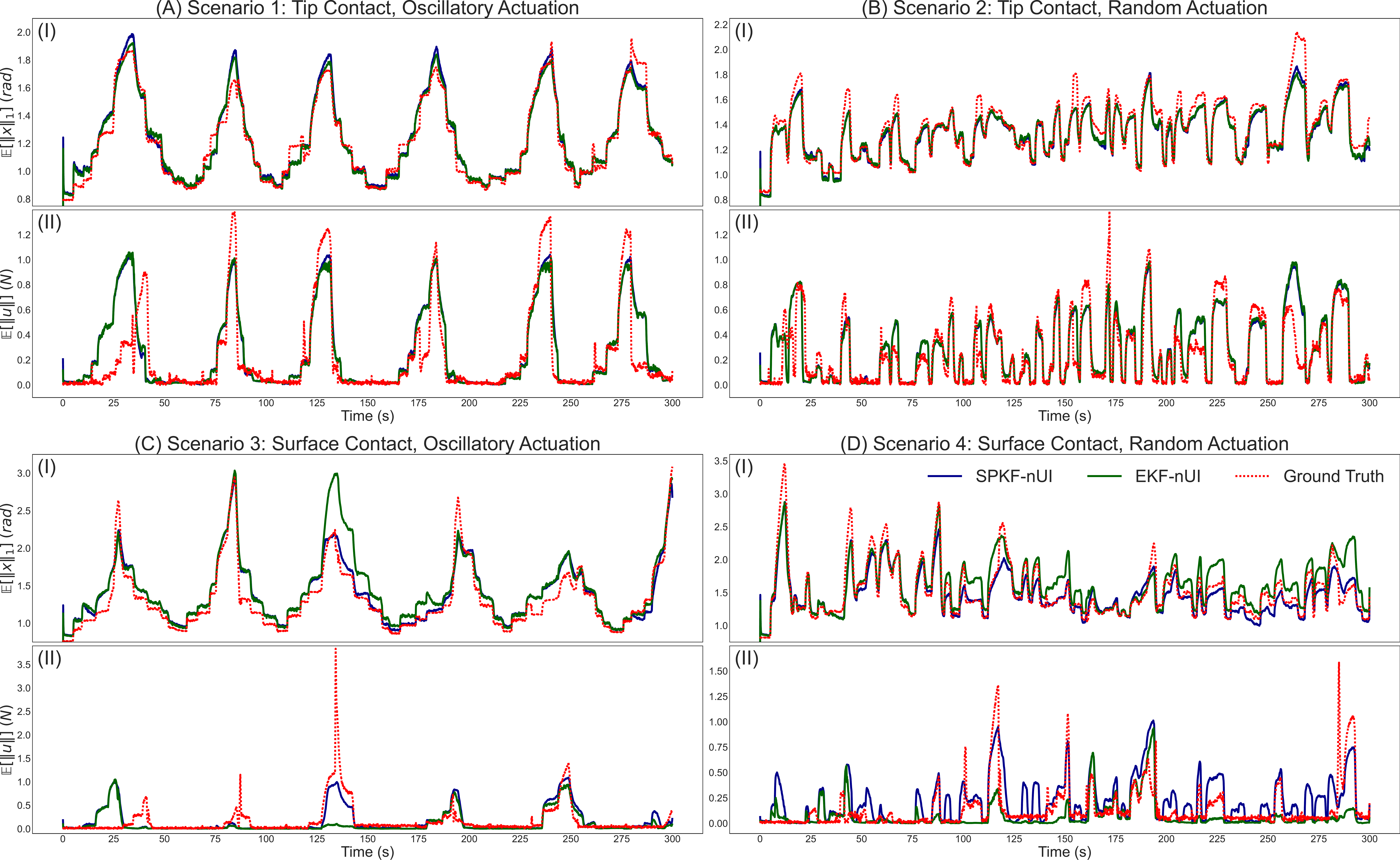}
\caption{\textbf{Estimation results of Case Study 2.}
The results of the four combinations: (\textit{Tip Contact}, \textit{Oscillatory Actuation}), (\textit{Tip Contact}, \textit{Random Actuation}), (\textit{Surface Contact}, \textit{Oscillatory Actuation}) and (\textit{Surface Contact}, \textit{Random Actuation}) are shown in (A), (B), (C), (D), respectively.
On each combination, 
(I) shows the ground truth $\|x_{t}\|_1$ and the mean state estimate $\mean[\|\hat{x}_{t}\|_1]$ of SPKF-nUI and EKF-nUI.
The 1-norm $\|x_{t}\|_1$ (physically) represents the aggregate bending angle.
(II) shows the ground truth $\|u_{t}\|$ and the mean UI estimate $\mean[\|\hat{u}_{t}\|]$ of SPKF-nUI and EKF-nUI.
The 2-norm $\|u_{t}\|$ (physically) represents the magnitude of the resultant contact force.
These results are obtained via averaging across the 10 MC simulations.
Legend in (B-I) applies to all plots.
}
\label{fig:SR_results}
\end{figure*}
\begin{table}[t]
\centering
\caption{\textbf{Overall State NMSEs $\mean[ \|\tilde{x}\|^2 ]$, UI NMSEs $\mean[ \|\tilde{u}\|^2 ]$ of Case Study 2.} NA indicates that result is not available. NC indicates non-converging (very large) errors.}
\label{tab:CaseStudy2}
\resizebox{\columnwidth}{!}{%
\begin{tabular}{@{}lcccc@{}}
\toprule
\multicolumn{1}{c}{\multirow{2}{*}{Method}} & \multicolumn{2}{c}{System State} & \multicolumn{2}{c}{Unknown Input} \\ \cmidrule(l){2-5} 
\multicolumn{1}{c}{} & \multicolumn{1}{c}{NMSE} & \multicolumn{1}{c}{SNR} & NMSE & \multicolumn{1}{c}{SNR} \\ \midrule
SPKF-nUI & \textbf{0.513 ± 0.023} & 43.5 ± 0.2 & \textbf{0.866 ± 0.020} & 26.1 ± 0.3 \\
EKF-nUI & 0.974 ± 0.055 & 40.7 ± 0.2 & 1.069 ± 0.033 & 26.4 ± 0.4 \\ \midrule
SPKF-nUI-I & 0.523 ± 0.026 & 43.5 ± 0.2 & 0.897 ± 0.024 & 25.9 ± 0.3 \\
EKF-nUI-I & 1.153 ± 0.131 & 40.6 ± 0.1 & 1.146 ± 0.058 & 26.7 ± 0.3 \\ \midrule
SPKF-nUI-II & 0.540 ± 0.045 & 42.7 ± 0.2 & 0.905 ± 0.030 & 26.3 ± 0.4 \\
EKF-nUI-II & 0.652 ± 0.085 & 42.5 ± 0.3 & 0.952 ± 0.057 & 25.8 ± 0.6 \\ \midrule
SPKF-MVU & 1.810 ± 0.006 & 46.4 ± 0.2 & 0.853 ± 0.003 & 22.5 ± 0.1 \\
EKF-MVU & \multicolumn{2}{c}{NC} & 0.854 ± 0.009 & 18.0 ± 0.3 \\ \midrule
SPKF-UI & 3.010 ± 0.021 & 44.5 ± 0.3 & 2.281 ± 0.023 & 21.6 ± 0.1 \\
EKF-UI & 0.706 ± 0.013 & 41.2 ± 0.2 & 1.044 ± 0.009 & 22.8 ± 0.1 \\ \midrule
SPKF & 0.566 ± 0.031 & 42.7 ± 0.1 & \multicolumn{2}{c}{NA} \\
EKF & 0.618 ± 0.014 & 42.1 ± 0.2 & \multicolumn{2}{c}{NA} \\ \bottomrule
\end{tabular}
}
\end{table}
\begin{table*}[t]
\centering
\caption{\textbf{Average processing time per time-step of case study 2.}}
\label{tab:SR_time}
\resizebox{\textwidth}{!}{%
\begin{tabular}{@{}lcccccccccccc@{}}
\toprule
Method & SPKF-nUI & EKF-nUI & SPKF-nUI-I & EKF-nUI-I & SPKF-nUI-II & EKF-nUI-II & SPKF-MVU & EKF-MVU & SPKF-UI & EKF-UI & SPKF & EKF \\ \midrule
Time Elapsed (s) & \multicolumn{1}{r}{0.0173} & \multicolumn{1}{r}{0.0145} & \multicolumn{1}{r}{0.0170} & \multicolumn{1}{r}{0.0141} & \multicolumn{1}{r}{0.0159} & \multicolumn{1}{r}{0.0144} & \multicolumn{1}{r}{0.0173} & \multicolumn{1}{r}{0.0267} & \multicolumn{1}{r}{0.0162} & \multicolumn{1}{r}{0.0137} & \multicolumn{1}{r}{0.0157} & \multicolumn{1}{r}{0.0130} \\ \bottomrule
\end{tabular}%
}
\end{table*}

Estimation results of the PSA case study are tabulated in Table \ref{tab:CaseStudy2}.
Statistical t-tests conducted over the 10 estimation samples at a significance level of 0.1 show that the overall (over four experimental scenarios) NMSEs of the SPKF-nUI are the lowest among the compared baseline filters.
It shows that our proposed SPKF-nUI achieves in overall the lowest state and UI NMSEs, in consistent with results obtained in the Case Study 1.
Fig. \ref{fig:SR_results} shows the estimations and the NMSEs of the SPKF-nUI and the EKF-nUI. 
It shows that the EKF-nUI performs worse than the proposed SPKF-nUI due to large remainder errors arising from linearization of the highly nonlinear NN models (\ref{eq:ss_GRU}).
In Table \ref{tab:CaseStudy2}, the SPKF-nUI-I and EKF-nUI-I that estimate UIs based on the less accurate prior state estimates perform significantly worse, compared to the SPKF-nUI, EKF-nUI which employ posterior state estimates.

Nonetheless, the NMSE advantages of the SPKF-nUI and EKF-nUI over their variants diminish in this case study, due to incorporation of the implicit recurrent hidden states $h_{t}$ in GRU.
Despite the linear least-squares UI optimization being replaced by the data-driven UI model (\ref{eq:ui_model}), SPKF-MVU, EKF-UI and most notably EKF-MVU, SPKF-UI perform poorly in state estimation due to the negligence of UI uncertainties and immoderate Kalman gain of these baselines.
Table \ref{tab:SR_time} tabulates the processing time of each filter for the PSF case study which is more involved. 
Compared to SPKF-nUI-II, SPKF-nUI requires more computational time to accommodate the proposed joint sigma-point transformation scheme (\ref{eq:sgms_posterior})-(\ref{eq:Pcovariance_predict}) for better robustness against UI uncertainties, which leads to an improvement in overall estimation accuracy.
SPKFs generally have higher computational time than EKFs, but it is far outweighed by the superior estimation performances. 
\section{Conclusion} \label{sect:Conclusion}
In this paper, we presented a derivative-free SPKF-nUI where the SPKF is interconnected with a general nonlinear UI estimation, performed via nonlinear optimization or data-driven approaches. The proposed method overcomes the common assumption of linearly separable UI in the system model.
Compared to existing approaches, the UI estimation of SPKF-nUI uses the posterior state estimate which is less susceptible to prediction errors.
In addition, the SPKF-nUI employs a sigma-point transformation scheme alongside UI estimation to incorporate the UI errors and uncertainties.
Furthermore, we conducted a stochastic stability analysis and proved that the SPKF-nUI yields exponentially bounded estimation errors. Lastly, we carried out two case studies to validate the efficacy SPKF-nUI, where results showed that it performs best among the existing filters. In conclusion, the proposed SPKF-nUI achieved accurate multi-modal state and UI estimations, crucial in realizing reliable perceptions for complex intelligent autonomous systems.
For future work, a robust filtering scheme for model disturbances exhibiting both deterministic and non-Gaussian characteristics could be considered. Recursive or incremental identification methods could also be explored to reduce the current dependency of learning-based UI model on prior data.
\bibliographystyle{IEEEtran}
\bibliography{ref}

\vspace{-0.75cm}
\begin{IEEEbiography}[{\includegraphics[width=1in,height=1.25in,clip,keepaspectratio]{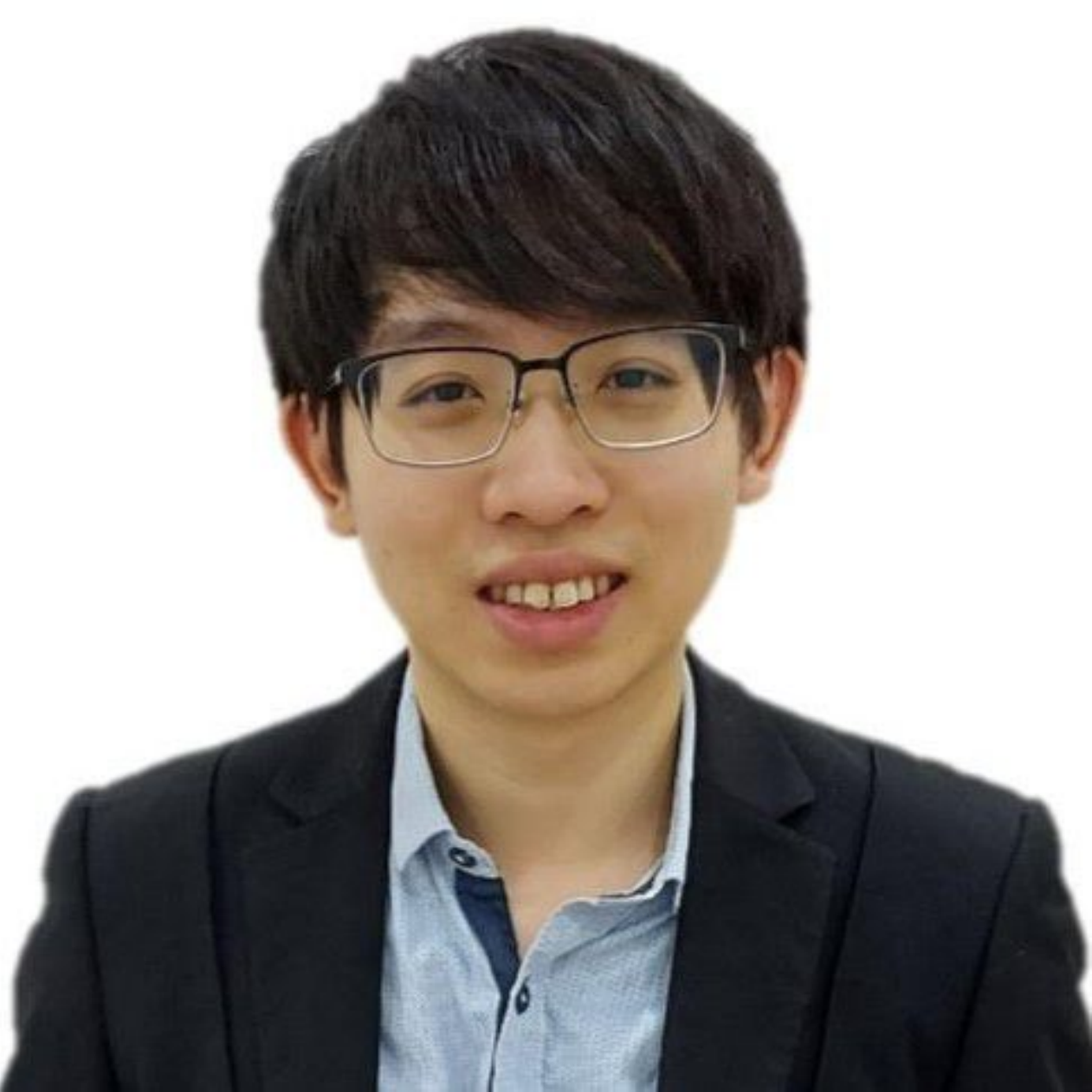}}]
{Loo Junn Yong} received his B.Eng (Hons) and Ph.D. degrees from the School of Engineering at Monash University Malaysia in 2018 and 2022, respectively. In the following year, he returned to Monash University Malaysia for a research fellowship until 2023. He is currently a Lecturer at the School of Information Technology, Monash University Malaysia. His research interests include generative modelling, nonlinear filtering, and intelligent robots.
\end{IEEEbiography}

\vspace{-0.75cm}
\begin{IEEEbiography}[{\includegraphics[width=1in,height=1.25in,clip,keepaspectratio]{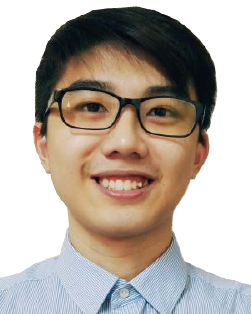}}]
{Ze Yang Ding} received the B.Eng. (Hons.) and Ph.D. degrees from Monash University Malaysia in 2019 and 2023 respectively. He is currently a Lecturer at the School of Engineering in Monash University Malaysia. His research interests include deep learning, data-driven modelling and soft sensors.
\end{IEEEbiography}

\vspace{-0.75cm}
\begin{IEEEbiography}[{\includegraphics[width=1in,height=1.25in,clip,keepaspectratio]{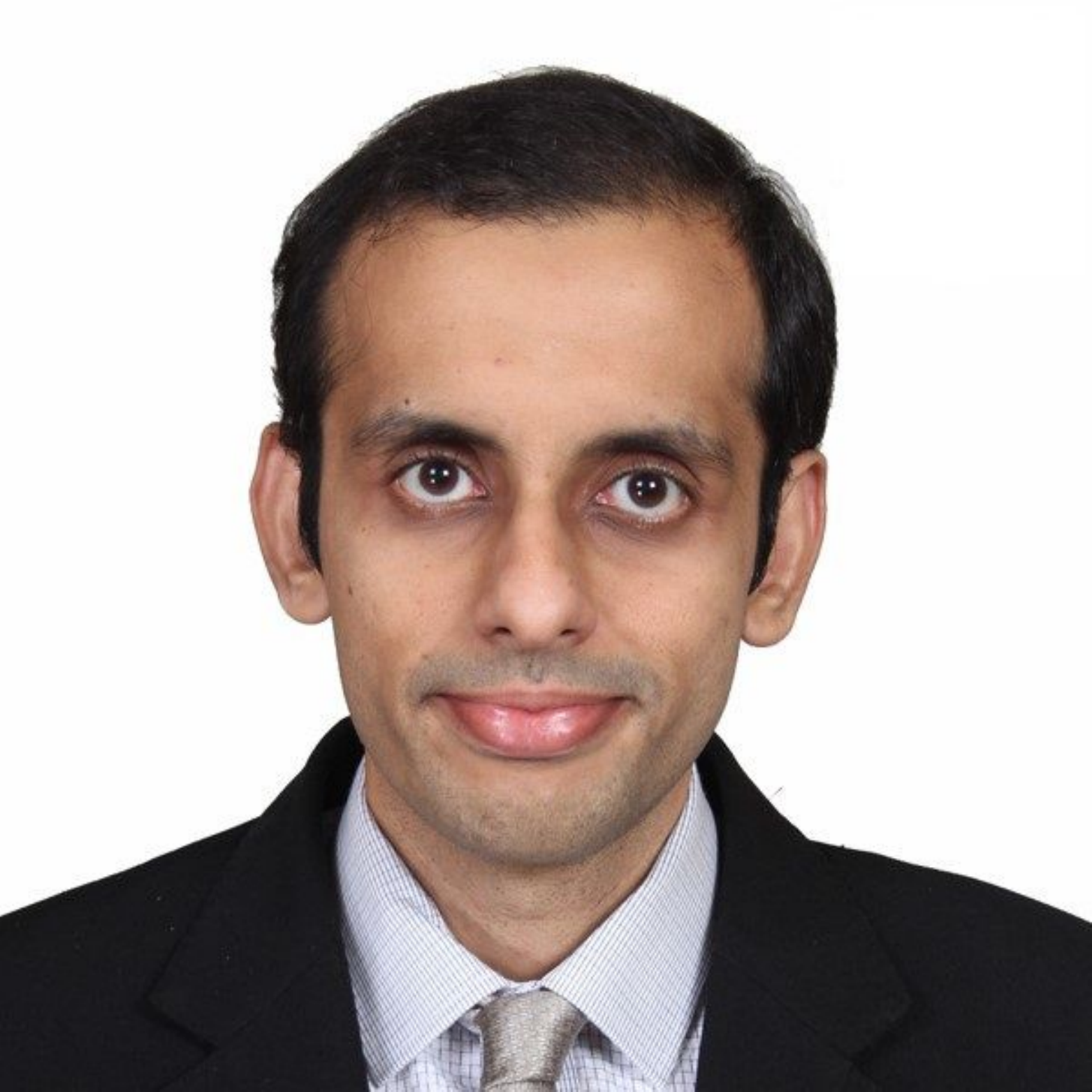}}]
{Vishnu Monn Baskaran} received his B.Eng. and M.Eng. degrees in Electrical and Electronics Engineering from Universiti Tenaga Nasional, Malaysia in 2004 and 2007, respectively. In 2016, he obtained his Ph.D. in Engineering from Multimedia University, Malaysia. His research interests include computer vision for object detection, human-to-object interaction, and time series deep learning for soft robot perception.
\end{IEEEbiography}

\vspace{-0.75cm}
\begin{IEEEbiography}[{\includegraphics[width=1in,height=1.25in,clip,keepaspectratio]{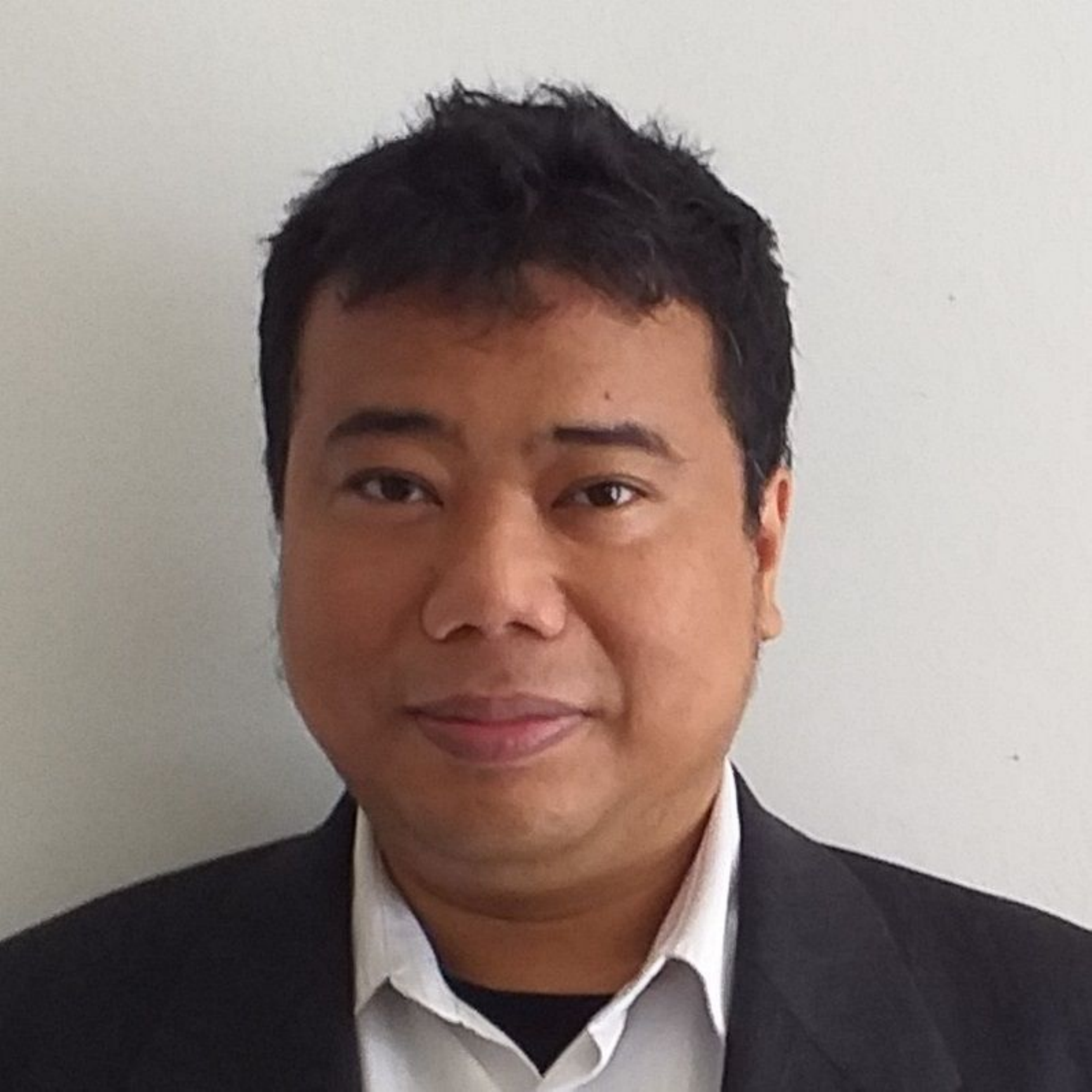}}]
{Surya G. Nurzaman} (Member, IEEE) received the Ph.D. degree in adaptive machine systems from Osaka University, Suita, Japan, in 2011, with the support of the Monbukagakusho scholarship from the Japanese government. He is a Senior Lecturer with the School of Engineering, Monash University Malaysia, Subang Jaya, Malaysia. From 2011 to 2015, he was a Research Fellow with Osaka University, ETH Zurich, Zurich, Switzerland, and the University of Cambridge, Cambridge, U.K. His research interests include soft-robotics, embodied intelligence, and machine learning. He’s a founding Editor-in-Chief of Robotics Reports (2023-now), as well as Associate Editor of Frontiers in Robotics and AI (2022-now) and IEEE Robotics and Automation Magazine (2021-2023). He's also an advisory board member of the IEEE RAS Technical Committee on Soft Robotics (2021-now).
\end{IEEEbiography}

\vspace{-0.75cm}
\begin{IEEEbiography}[{\includegraphics[width=1in,height=1.25in,clip,keepaspectratio]{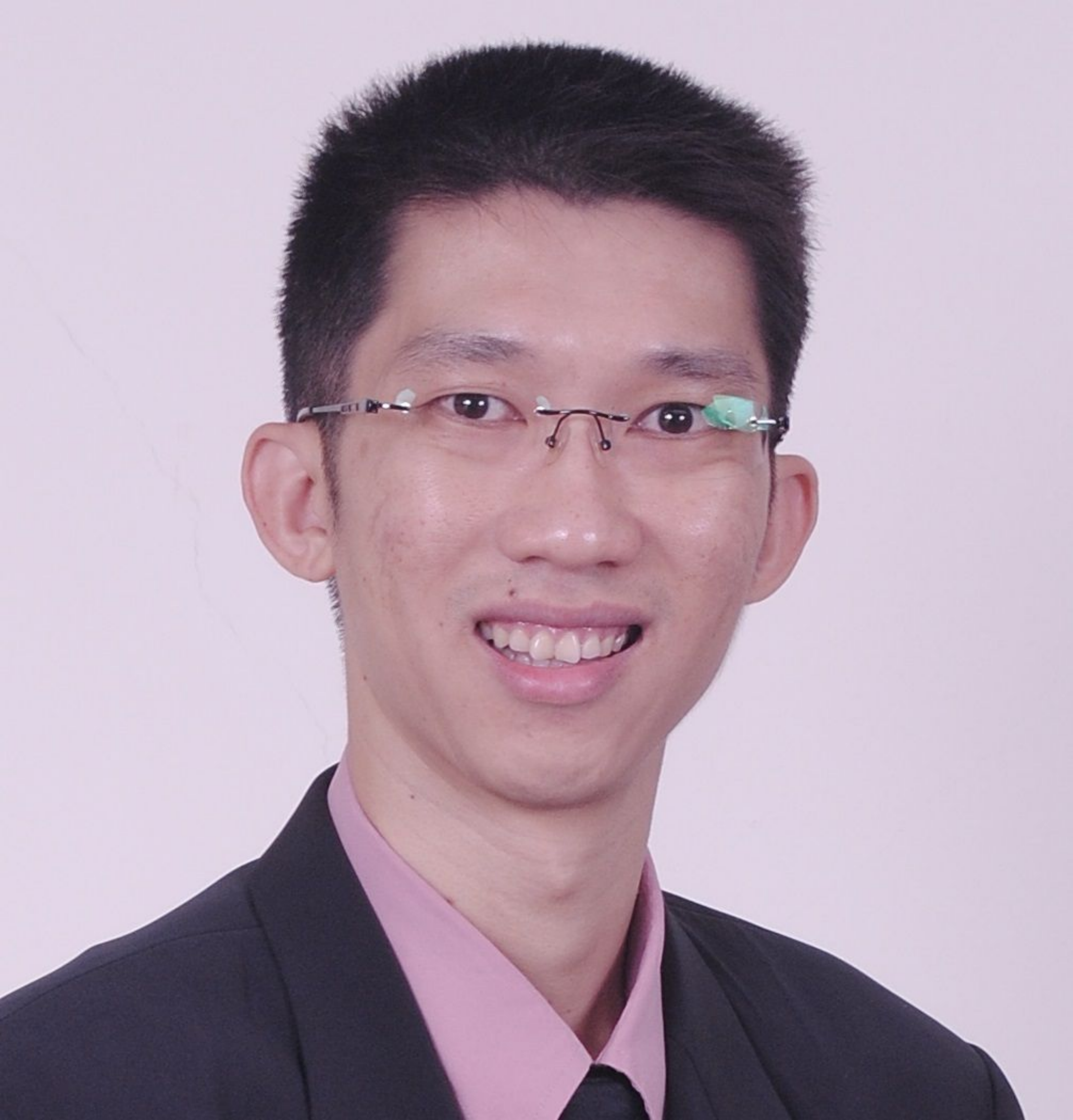}}]
{Chee Pin Tan} (Senior Member, IEEE) received the B.Eng. (Hons.) and Ph.D. degrees from University of Leicester, Leicester, U.K., in 1998 and 2002, respectively. He is currently a Professor with the School of Engineering, Monash University Malaysia, Bandar Sunway, Malaysia. He has authored more than 80 internationally peer-reviewed research articles, including a book on fault reconstruction. His research interests include robust fault estimation and observers. He is currently an Associate Editor for IEEE Transactions on Cybernetics, Journal of the Franklin Institute, and International Journal of Systems Science.
\end{IEEEbiography}

\end{document}